\documentclass[12pt]{iopart}
\pdfoutput=1

\usepackage{iopams} 
\usepackage{graphicx}
\usepackage{color}
\usepackage{bm}
\usepackage{amssymb}
\usepackage{hyperref}
\usepackage{soul}
\usepackage{cite}

\expandafter\let\csname equation*\endcsname\relax

\expandafter\let\csname endequation*\endcsname\relax

\usepackage{amsmath}

\usepackage{todonotes}
\presetkeys{todonotes}{size=\footnotesize, color=yellow!90}{}

\begin{document}

%\title{Controlling the energy gain-loss balance in stochastic oscillators} 
%\title{{\cal PT} symmetry and phase transition for characterizing 
%the energy gain-loss balance in stochastic oscillators} 
\title{Resilience of $\mathcal{PT}$ symmetry against stochasticity 
in a gain-loss balanced oscillator}

\author{Mirko Lukovi\'c$^{1}$\footnote{Author to whom any correspondence should be addressed.
}, Patrick Navez$^2$, Giorgos P Tsironis$^2$ and Theo Geisel$^{3}$}

\address{$^1$ Institute for Building Materials, ETH Zurich, 8092 Zurich, Switzerland}
\address{$^2$ Department of Physics, University of Crete, P. O. Box 2208, 71003 Heraklion, Greece}
\address{$^3$ Max Planck Institute for Dynamics and Self-Organization, 37077 G\"{o}ttingen, Germany}
\ead{lukovicm@ethz.ch}

\vspace{10pt}
\begin{indented}
\item[]August 2016
\end{indented}

\begin{abstract}
We investigate the effects of dichotomous noise added to a classical harmonic oscillator in the form of stochastic time-dependent gain and loss states, whose durations are sampled from two distinct exponential waiting time distributions. Despite the stochasticity, stability criteria can be formulated when averaging over many realizations in the asymptotic time limit and  serve to determine the boundary line in parameter space that separates regions of growing amplitudes from those of decaying ones. Furthermore, the concept of $\mathcal{PT}$ symmetry remains applicable for such a stochastic oscillator and we use it to distinguish between an underdamped symmetric phase and an overdamped asymmetric phase. In the former case, the limit of stability is marked by the same average duration for the gain and loss states, whilst in the the latter case, a higher duration of the loss state is necessary to keep the system stable. The overdamped phase has an ordered structure imposing a position-velocity ratio locking and is viewed as a phase transition from the underdamped phase, which instead displays a broad and more disordered, but nevertheless, $\mathcal{PT}$ symmetric structure. We also address the short time limit and the dynamics of the moments  of the position and the velocity with the aim of revealing the extremely rich dynamics offered by this apparently quite simple mechanical system. The notions established so far may be extended and applied in the stabilization of light propagation in metamaterials and optical fibres with randomly distributed regions of asymmetric active and passive media.

\end{abstract}

%\todo{PACS, keywords}
\pacs{05.40.-a, 05.70.Fh, 11.30.Er, 42.25.Dd}
%11.30.Er, 42.25.Bs, 42.82.
%\pacs{03.65.Ge, 03.65.Sq, 05.40.-a, 03.65.Vf,05.70.Fh}
\vspace{2pc}
\noindent{\it Keywords}: Stochastic oscillators, Dichotomous noise, PT-symmetry, Phase transition. 

\submitto{\NJP}
%\maketitle

\section{Introduction}

%\todo{In the intro, we should separate well the work of the other 
%from what we are doing}

Given that physical systems are in general not conservative but rather tend to dissipate energy, some external forces are always necessary in order to reactivate their dynamics. There exist many systems, such as the simple pendulum, a motor engine or electric circuit that undergo transitions between states in which energy is gained and dissipated. If the gain is tunable enough in order to compensate the loss, then the resulting device simulates perfectly a conservative system and thus preserves the time reversal symmetry ${\mathcal T}$ but for many reasons, mostly technical (regulation or automatism), the compensation is not always perfect so that the loss and gain have to be treated separately.

However, even an imperfect control of the energy balance can result in extraordinary new properties. Indeed, let us consider a system that switches between two possible states, one in which energy is gained and the other where energy is lost. It is clear that such a system breaks the time reversal invariance ${\mathcal T}$ and looses its nice property of energy conservation. What is more obvious is that it is not invariant under the operation ${\mathcal P}$ which consists in a swap between the loss and gain states. Yet, in order to preserve some symmetry, one idea is to combine the two operations and impose on the non-conservative system a weaker requirement of invariance under the $\mathcal{PT}$ transformation.

Works involving $\mathcal{PT}$ symmetry have been initiated in the context of quantum mechanics using non Hermitian Hamiltonians \cite{bender1998real} in which ${\mathcal P}$ refers to the parity operator, which reverses the position. The investigation of a generic class of $\mathcal{PT}$ symmetric Hamiltonians has shown that their energy spectrum remains real below a critical point but becomes imaginary beyond it, manifesting a transition to new 'exotic' quantum states. Subsequent works \cite{bender1998real,bender2002generalized,bender2003must,bender2007making} have led to a reformulation of the use of these concepts in the framework of a non-Hermitian model involving only two quantum states for which the parity operator corresponds to the swap operation between the two states.

Since these seminal works \cite{bender1998real,bender2002generalized,makris2008beamDynamics,christo2009complexPotentials,ruter2010observation}, this new field has emerged in other contexts such as classical optics and electrical circuits in order to better understand the interplay between active and passive transmission, but also in tight binding systems \cite{longhi2013tightBinding,joglekar2010robust,joglekar2011robust}. In optical fibers, the simultaneous use of active and passive components  displays very interesting properties such as transient wave amplification in an array of coupled waveguides with an arbitrary space distribution of gain and loss \cite{makris2014anomalous}. Furthermore, there are experiments which demonstrated that $\mathcal{PT}$ symmetric materials can exhibit power oscillations, non-reciprocal light propagation and tailored energy flow \cite{christo2009complexPotentials,ruter2010observation,schindler2011experimental}. In addition, the existence of giant amplifications is predicted, meaning that a passive medium may be helpful to enhance the gain effect of an active medium \cite{konotop2012PT}. Similar problems were studied in another experiment with a pair of coupled oscillators in the form of an $LRC$ circuit \cite{schindler2011experimental}. Instead of considering a single oscillator that switches between gain and loss states, the authors of \cite{schindler2011experimental} examined an electronic dimer made of two coupled oscillators, one with gain and the other with loss. The experiment succeeds in displaying all the phenomena encountered in systems with generalized $\mathcal{PT}$-symmetries. 

In order to understand the basics of a $\mathcal{PT}$-symmetric gain and loss process, a very simple one-dimensional harmonic oscillator was considered. The prototype model consisted of two separate states of frictional and gain forces linearly proportional to the velocity that alternate periodically in time \cite{tsironis2014ptZeroD}. 
%In a recent publication, Lazarides and Tsironis considered such a model with a time dependent damping coefficient \cite{tsironis2014ptZeroD}. 
%Consequently, at regular time intervals the oscillator was made to pass from %a damping state to an anti-damping state.  
%The authors demonstrated that for carefully chosen time intervals, such a %classical system obeys the combined parity-time ($\mathcal{PT}$) symmetry %\cite{bender1998real,ruter2010observation,schindler2011experimental}. 
%[{\color{blue}\emph{Here it should be explained in very simple and intuitive %terms what exactly $\mathcal{PT}$-symmetry is. It should also be explained %with an example in optics why PT-symmetry is important and what new %properties are obtained from it.}}]
This model contains only the oscillator frequency, the damping coefficient and the alternating period as parameters. Quite remarkably, it provides a complete analysis with a phase diagram that distinguishes the stable from the  unstable regimes according to the parameter values.

%We, on the other hand, for the first time
However, as the dynamics might be even less controllable in the presence of randomness, a natural question arises on how it affects, or rather breaks the $\mathcal{PT}$-symmetries. In this paper, we investigate how an effective ${\mathcal PT}$ symmetry persists in the presence of dichotomous noise introduced by replacing the fixed time periods with random intervals in the simple generic model developed in \cite{tsironis2014ptZeroD}. More precisely, the oscillator switches randomly in time between a damping state in which energy is dissipated or lost and an anti-damping state in which energy is accumulated or gained. This oscillator can represent one electromagnetic mode in a cavity that is amplified randomly
in order to compensate the losses.

%study the effects of dichotomous noise in the context of a two state system %with a gain-loss interplay. 
%\st{As the simplest model/prototype that captures the gain-loss feature, we %consider a one-dimensional harmonic oscillator with a damping coefficient that can be made to switch between a positive and negative value. The oscillator essentially switches between a damping state in which energy is dissipated or lost and an anti-damping state in which energy is accumulated or gained.}

There exist earlier studies of the effects of random damping on the stability of harmonic oscillators \cite{gitterman2014pre,mendez2011instabilities}. The stabilities of the first two moments of the oscillator position and velocity have been analyzed, but only for uncorrelated Gaussian and colored noise and not for dichotomous noise. In this context, we also mention the work on the ${\mathcal PT}$ symmetric coupler in \cite{konotop2014stochastic} with Gaussian white noise, where amplification occurs despite the perfect balance of gain and loss. In contrast to these previous works, besides  determining the moments, we are also able to characterize in the asymptotic limit the exact nature of the probability distribution generated by the random noise and thus predict the oscillator energy distribution. Furthermore, we also introduce an alternative notion of stability based on the energy logarithm of the system which we motivate through the properties of the probability density function of the state of the system. In addition to the ${\mathcal PT}$ symmetric states, we also found regimes in which this symmetry is broken even though the average durations for the loss and gain states were equal. This observation confirms the known statement that energy amplification occurs even when the system is predominantly dissipative over time 
\cite{makris2014anomalous, konotop2014stochastic} and can be formally established using a mathematical framework based on the master equation. Finally, we succeed in pointing out the analogy with phase transitions in thermodynamics, in which beyond a certain critical value, the stochastic oscillator breaks its ${\mathcal PT}$ symmetry towards an ordered phase.

%We go further and introduce a randomly fluctuating damping coefficient so %that a single oscillator passes through consecutive gain and loss phases at %random time intervals. Such stochasticity in the system breaks time reversal %and therefore also the possibility of having $\mathcal{PT}$-symmetry.

%[{\color{blue} \emph{I guess this paragraph can be improved. The way it is written it looks like the earlier work diminishes our results.}}]
%Finally, we relate the symmetries of the system to its stability through the study of the probability density functions of the state of the system.
%Finally, we show analytically using a mathematical framework based on the master equation, that it is possible to have average energy amplification even when the system is predominantly dissipative over time. Although the evolution of the system is stochastic and irreversible, it is possible \st{to recover} nevertheless to provide a definition of $\mathcal{PT}$-symmetry in the sense of an average over an ensemble of stochastic oscillators. 

%\begin{itemize}
%\item when we talk about an oscillator, we talk about modes and structures. You can mention one of Makris' works. Randomness affects the stability diagram.
%\end{itemize}

This paper is organized as follows. In section 2, we formulate the stochastic oscillator problem in terms of the master equation and define an asymptotic stability criterion. In section 3, we present the results for both simulations and analytics and show how they can be related to a phase transition. Section 4 concerns a more restricted stability criterion involving  the position and velocity averages. We discuss the short time behaviour and stability involving higher order moments of the velocity in the limit of zero frequency oscillation in section 5 before ending with the conclusion in section 6.

\section{Stochastic harmonic oscillator}

\subsection{General consideration}

\begin{figure}[ht]
\begin{center}
\includegraphics[scale=0.55]{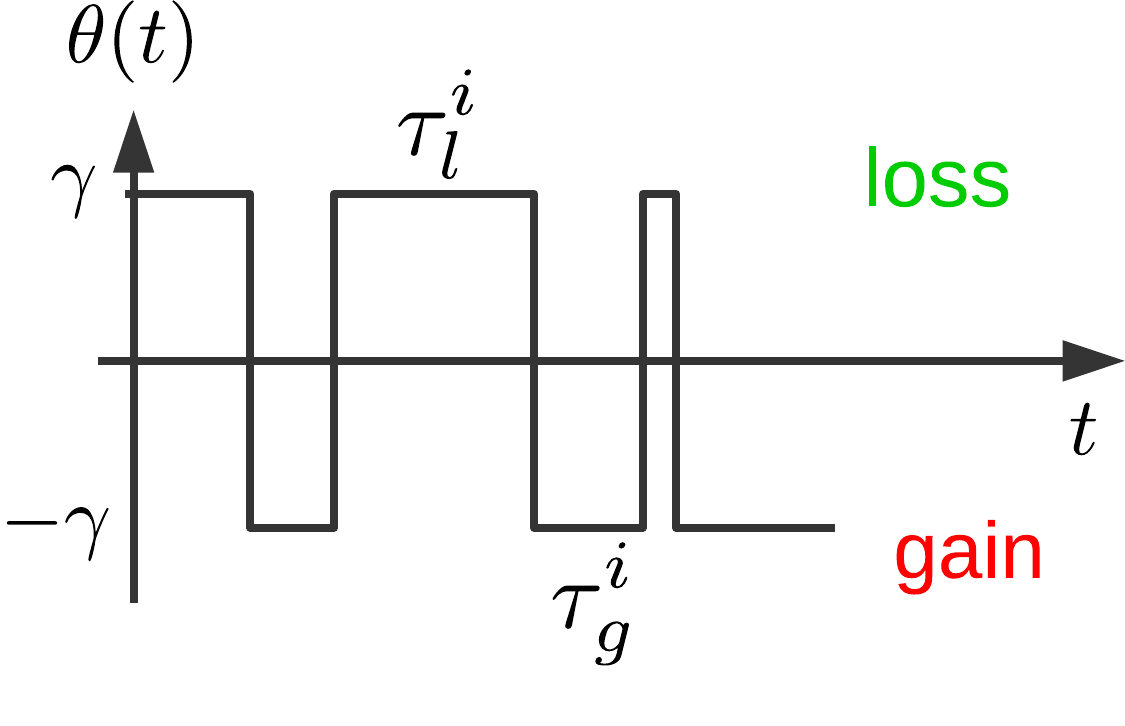} 
\hspace{2cm}\includegraphics[scale=0.55]{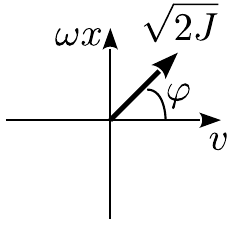}
\end{center}
\caption{LEFT: The time dependence of the damping function $\theta(t)$. When $\theta=\gamma$ the oscillator undergoes damping (loss) and when $\theta=-\gamma$, the oscillator undergoes anti-damping (gain). The amount of time spent in the gain/loss state before switching is $\tau_{g/l}$. RIGHT: Change of variables.}
\label{fig:theta}
\end{figure}

We consider a simple harmonic oscillator that randomly switches between a damping and anti-damping phase. The equation of motion of such an oscillator with natural angular frequency $\omega$ has the form
\begin{equation}\label{eq:damped_harm_osc}
\ddot{x}+2\theta(t)\dot x + \omega^2 x = 0,
\end{equation}
with a time-dependent damping coefficient $\theta(t)$ that can take only two constant values, either $+\gamma$ or $-\gamma$, i.e. $\theta(t)$ is a piece-wise constant function in the form of dichotomous noise (see figure \ref{fig:theta}). In the former case, the oscillator undergoes damping and therefore loses energy (loss state) whilst in the latter case it gains energy (gain state).  
We introduce stochasticity through the damping coefficient so that the system fluctuates between the gain ($g$) and loss ($l$) phases with residence times $\tau$. For the gain or loss phase the residence time is sampled from a distinct exponential distribution of the form $\tau_{g/l}^{-1}\exp(-\tau/\tau_{g/l})$. The two phases are therefore characterized by well defined average residence times, $\tau_g$ in the case of gain and $\tau_l$ in the case of loss.

\subsection{Master equation}

%When dealing with a stochastic system, the natural starting point is to try and determine 
In order to deal with the stochastic system, we begin by examining the time evolution of $p(x,v, t)$, the probability to find the oscillator in position $x$ with velocity $v=\dot x$ at time $t$. To this end we write down the master equation of such a system, keeping in mind that the oscillator can also be in any of the two states $g$ or $l$. Therefore, we have the following system of coupled partial differential equations:
\begin{multline}\label{eq:ME}
[\partial_t + v\partial_x - \omega^2 x \partial_{v}]p_{g/l}(x,v,t) \pm 2\gamma \partial_{v} (v p_{g/l}(x,v,t))=\\
\mp (p_g(x,v,t)/\tau_g-p_l(x,v,t)/\tau_l),
\end{multline}
such that $P_{g/l}(t)=\int_{-\infty}^\infty\int_{-\infty}^\infty dx\;dv\;p_{g/l}(x,v,t)$ is the probability for the oscillator to be in the gain/loss state at time $t$. Furthermore, the  probability is conserved so that $P_g(t)+P_l(t)=1$. The first term on the left-hand-side of (\ref{eq:ME}) is the deterministic Liouville term, whilst the second one is the gain/loss term that arises from the non-conservative nature of equation (\ref{eq:damped_harm_osc}). The term on the right-hand-side describes the stochastic switching rate of the oscillator between the gain and loss phases. 

%It is important to note that the two rates $1/\tau_l$ and $1/\tau_g$ in (\ref{eq:ME}) do not actually depend on the form of the waiting time distribution used. In that sense, the master equation is more general. 

In as such, one would have to solve the coupled pair of equations in (\ref{eq:ME}) for $p_g(x,v,t)$ and $p_l(x,v,t)$ in order to completely characterize the stochastic system. However, such a task is extremely heavy
%difficult 
and unnecessary for our purpose. 
As mentioned in the introduction, we are interested in the stability of the stochastic oscillator and a reliable criterion for it. 
A first simplification arises by noticing  that the action-angle
variables 
\begin{equation}
J=\frac{1}{2}\left(v^2+\omega^2 x^2\right)\quad\text{and}\quad\varphi = \arctan(\omega x/v),
\end{equation}
are more suitable for handling the master equation.
%In what follows, we show that it is possible to find a simple expression for %the evolution of the average logarithm of the energy and that it is via this %quantity that we can determine whether the system is stable or not. 
These polar-type coordinates lead to a separation of variables in the master equation in (\ref{eq:ME}) (see \ref{sec:unperturbed_problem} for details). In the optics terminology, $J$ and $\varphi$ correspond respectively to the amplitude and phase while $x$ and $v$ correspond to the quadrature components.
%[{\color{bf lue} \emph{Should be explained better}}]. Therefore, we apply the %following change of variables (see figure \ref{fig:theta}):

\begin{figure}[ht]
\begin{center}
\includegraphics[scale=1]{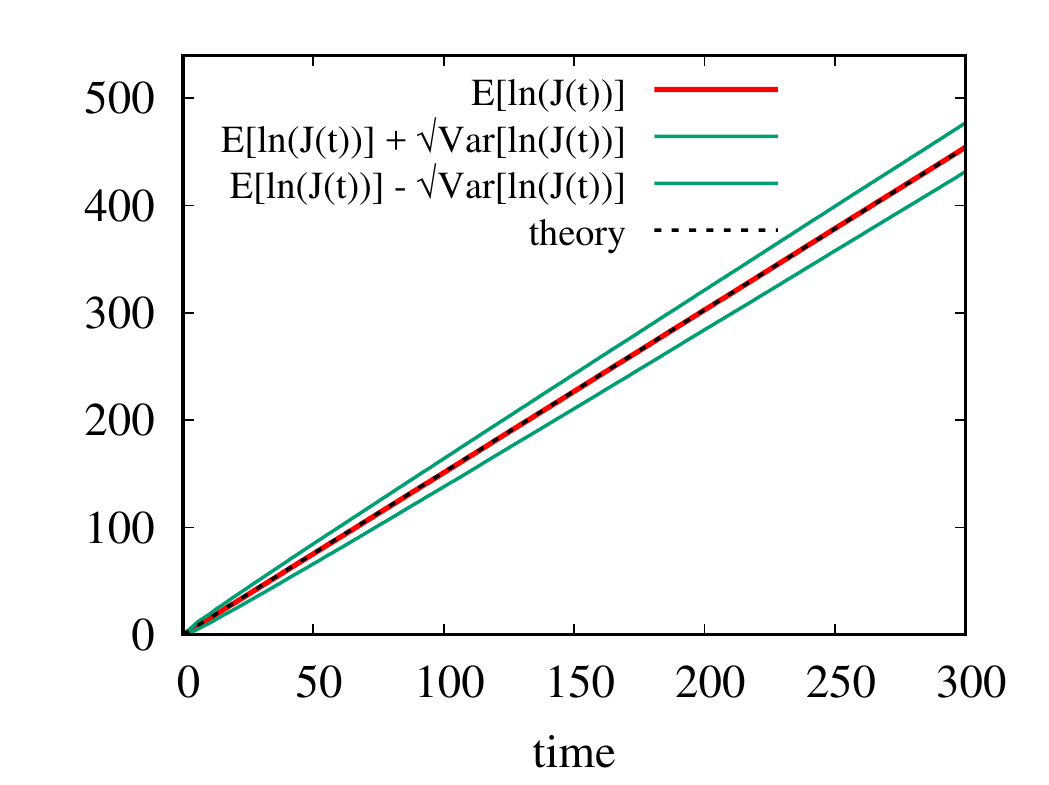} 
\end{center}
\caption{\textbf{Time evolution of the average of $\bm{\ln J(t)}$ together with its dispersion.} In this particular example $\omega=1$, $\gamma = 0.9$, $\tau_g=5$ and $\tau_l=1$. The average is taken over an ensemble of $10^4$ oscillators, each with the initial condition $(x_0,v_0)=(1,1)$ at $t=0$. The positive slope of $\ln J(t)$ for those particular parameter values implies that the system is unstable in the asymptotic limit. The dashed black line, whose slope was obtained from (\ref{eq:criterion}), corresponds to the theoretical result.}
\label{fig:variance}
\end{figure}

\begin{figure}[ht]
\begin{center}
\includegraphics[scale=1]{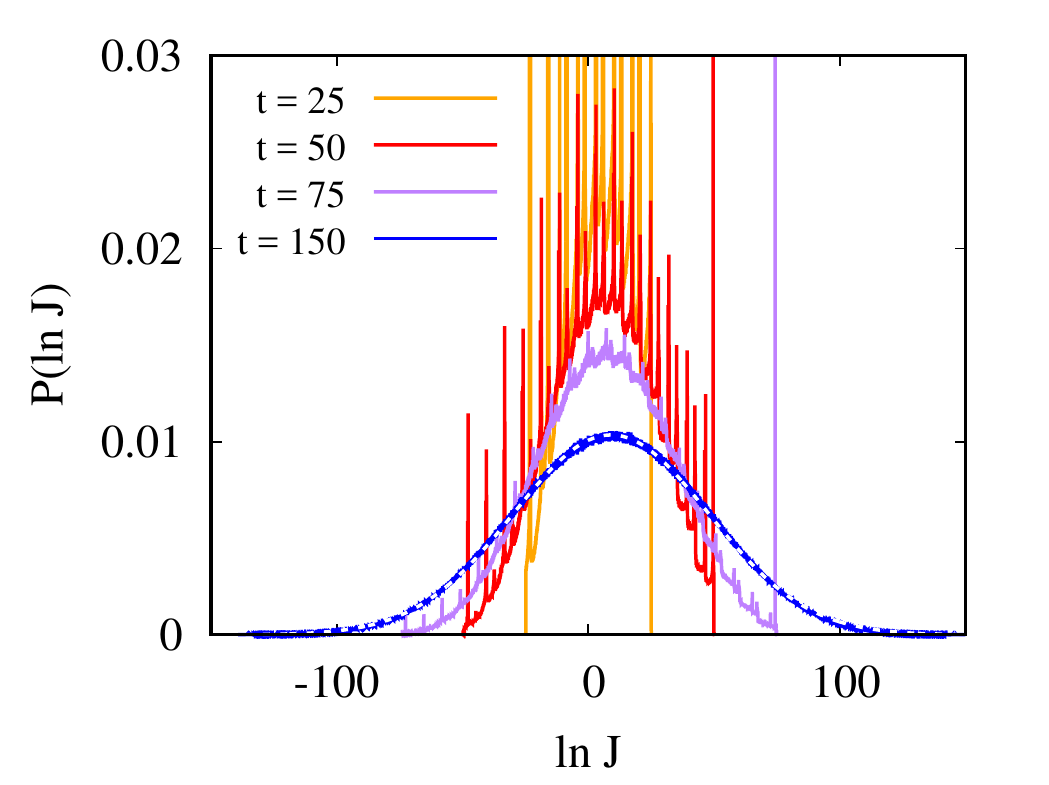} 
\end{center}
\caption{\textbf{Probability distribution of $\bm{\ln J(t)}$ for $\bm{\tau_g=\tau_l}$.} The different curves show the evolution of the probability distribution over different times $t$. For all the examples shown in the figure, $\omega=1$, $\gamma=0.5$ and $\tau_{g/l}=10$. The white dashed line is the Gaussian fit to the simulation result in blue. The fit parameters correspond to $\mu$ = 9.1 for the average and $\sigma$ = 38.8 for the standard deviation . An ensemble of $10^8$ oscillators was used to create both distributions, each oscillator having the initial condition $(x_0,v_0)=(1,1)$ at $t=0$.}
\label{fig:logJdistro}
\end{figure}

%[{\color{blue}\emph{write something regarding the choice of the new variables and how we decide to adopt those particular ones}}]:
%\begin{equation}
%J=\frac{1}{2}\left(v^2+\omega^2 x^2\right)\quad\text{and}\quad\varphi = %\arctan(\omega x/v),
%\end{equation}
Subsequent Laplace and Mellin transforms allow us to eliminate the time and $J$
derivatives in the resulting  master equation. Indeed, 
these transformations defined respectively as
%with respect to the $t$ and $J$ variables defined respectively as 
\begin{equation}\label{eq:transforms}
\hat f(s) = \int_0^\infty e^{-st}f(t)\; dt \quad\text{and}\quad \hat f(k) = \int_0^\infty J^k f(J)\; dJ.
\end{equation}
simplify the master equation into:
\begin{multline}\label{eq:ME_transform}
\left[\frac{d}{d\varphi}(\omega\mp\gamma\sin2\varphi) + s \mp 4\gamma k\cos^2\varphi\right]\hat p_{g/l}(k,\varphi,s) =\\ \mp \left(\frac{\hat p_{g/l}(k,\varphi,s)}{\tau_g} - \frac{\hat p_{g/l}(k,\varphi,s)}{\tau_l}\right) + \hat p_{g/l}(k,\varphi,t_0),
\end{multline}
where $p_{g/l}(J,\varphi,t_0)$ is the starting distribution (initial condition), which we assume to be a delta function. 
Still however, the last form cannot be solved analytically exactly. Nevertheless, essential information can be derived about the asymptotic time limit of the solution (see \ref{sec:dominant_eigenvalue} and \ref{sec:perturbation_theory}). For large times, we deduce indeed that the variable $\ln J$ follows a
normal distribution by showing that any moment of the cumulant expansion of $\ln J$ scales linearly with time  (see \ref{sec:evolution_moments}). As a consequence, the  average 
value $\langle \ln  J \rangle$
%, where $f(\phi)$ is an arbitrary function, 
depends linearly on time  and the relative square root 
variance has the scaling 
$\sqrt{\langle \delta^2 \ln J \rangle} /\langle 
\ln J \rangle \rightarrow 1/\sqrt{t}$ so that asymptotically $\ln J$ becomes a deterministic variable (see \ref{sec:criterion_appendix}). On the other hand, the angle $\varphi$ remains generally distributed over a broad value range. The numerical simulations of the evolution of an ensemble of stochastic oscillators confirm these  theoretical results: Figure \ref{fig:variance} shows the linear time dependence for the average around which the square root variance remains small in comparison; figure \ref{fig:logJdistro} shows the histogram of the distribution of $\ln J$, which converges to a Gaussian in the asymptotic time limit.

\subsection{The stability criterion}\label{sec:stability_criterion}

In order to assess the stability of the dynamics of the stochastic
oscillator, we can use the following results established in 
\ref{sec:criterion_appendix}.
If the asymptotic marginal distribution $P_{g/l}(\varphi)=\lim_{t\rightarrow\infty}\int_0^\infty p_{g/l}(J,\varphi,t)\;dJ$ exists, 
%the equation in (\ref{eq:criterion_fourier}) can be rewritten without the %transformed functions to produce the more intuitive expression
then the asymptotic constant $\eta$ associated to the linear evolution of the first moment is given by:
\begin{equation}\label{eq:criterion}
\eta=\lim_{t\rightarrow\infty}\frac{d}{dt}\langle \ln J \rangle(t) = \frac{\int_{-\pi/2}^{\pi/2}4\gamma\cos^2\varphi\;\left(P_g(\varphi)- P_l(\varphi)\right)d\varphi}{\int_{-\pi/2}^{\pi/2}\left(P_g(\varphi) + P_l(\varphi)\right)d\varphi},
\end{equation}
This real-valued constant is the basis of the stability criterion that we shall employ in the next section.
%to characterize the evolution of the stochastic oscillator in the asymptotic %limit. 
%where $\eta$ is a real-valued constant which we will use as the discriminant that determines the stability of the system.
A positive $\eta$ corresponds to a diverging first moment of $\ln J$ implying that the system is unstable while a negative $\eta$ corresponds to a stable system. %
%In the following section we will discuss in detail the stability of the %stochastic oscillator using (\ref{eq:criterion}).
In order to apply the stability criterion defined in  (\ref{eq:criterion}), 
we need to determine $P_g(\varphi)$ and $P_l(\varphi)$ asymptotically.

\section{Stability results of the stochastic oscillator in the asymptotic limit}

\subsection{Simulated results compared with the theory}

\begin{figure}
\begin{center}\label{fig:stability_sim}
\includegraphics[width=1\linewidth]{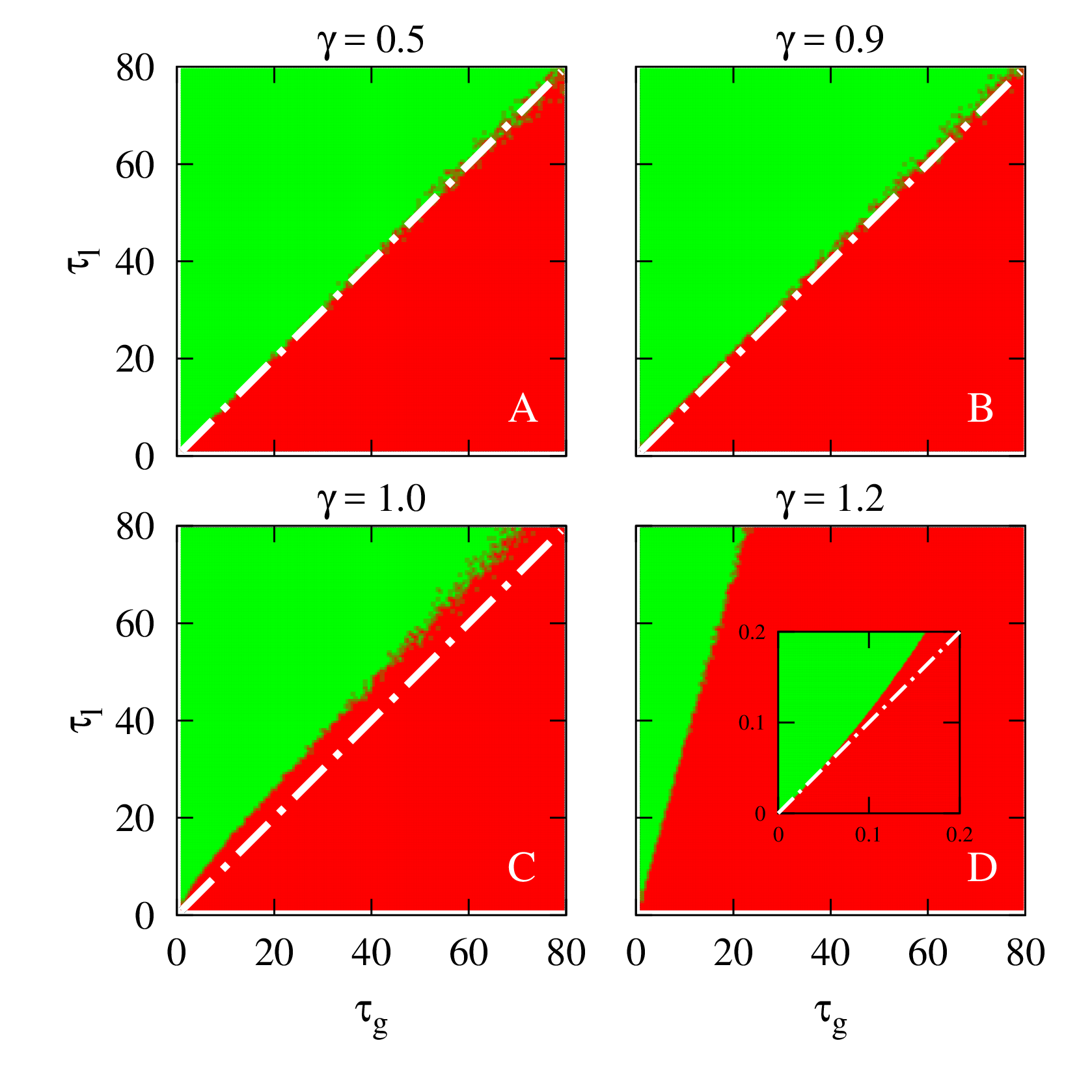}
\end{center}
\caption{\textbf{Stability of the stochastic oscillator.} The figure shows the stability analysis done using numerical simulations according to the criterion defined in (\ref{eq:criterion}). The green color indicates the stable region, where the average of $\ln J$  is negative  and the red color indicates the unstable region, where it is positive. \textbf{Panels A} and \textbf{B}: In this case $\gamma < \omega$. It can be seen that the line of symmetry $\tau_g=\tau_l$ separates the stable region from the unstable. \textbf{Panel C}: This is the critical case where $\omega=\gamma$. Since the system is not in the asymptotic regime, the symmetry line $\tau_g=\tau_l$ does not separate the two regions perfectly. \textbf{Panel D}: An example of the case where $\gamma > \omega$. It shows that the symmetry line $\tau_g=\tau_l$ is well below the line that separates the two regions. Consequently there exist cases where $\tau_g$ is well below $\tau_l$ and yet the system is still unstable. The \textbf{inset} shows that for small values of $\tau_g$ and $\tau_l$ symmetry is regained, as discussed at the end of \ref{sec:asymptotic_approx}. In all four cases an ensemble of 1000 oscillators that evolved up to $t=300$ were used, each oscillator having the initial condition $(x_0,v_0)=(1,1)$ at $t=0$.}
\label{fig:stability2}
\end{figure}

%In order to apply the stability criterion defined in  (\ref{eq:criterion}), 
%we need to determine $P_g(\varphi)$ and $P_l(\varphi)$ for a time much larger than 
%which, on the other hand, is not at all trivial. 
%We proceed here with an asymptotic approximation of (\ref{eq:criterion}) 
%in the limit where 
%$\tau_g,\tau_l\gg1$. 
We used numerical simulations of the evolution of an ensemble of stochastic oscillators from which we extract the asymptotic first moment of $\ln J(t)$ and thus determine $\eta$. The results are plotted in figure \ref{fig:stability2}. The green area corresponds to negative values of the slope of the ensemble average of $\ln J$ whilst the red corresponds to positive ones. For $\gamma<\omega$, the two regions are separated by the line of symmetry $\tau_g=\tau_l$. This result corresponds to what one might expect  - if the amount of time spent in the gain state is on average longer than in the loss state, then the average value of the energy diverges over time. On the other hand, if the system on average spends more time in a loss state, then its average energy  decays to zero over time. What comes as a surprise is that when $\gamma > \omega$ the system energy can diverge even when $\tau_l > \tau_g$. In order to see this better it is worth theoretically studying the dependence of $\eta$ in terms of $\gamma$.

In the limit of large $\tau_g$ and $\tau_l$, we calculate explicitly the  formula (\ref{eq:criterion}) using the expressions for $P_{g/l}(\varphi)$ 
derived in  \ref{sec:unperturbed_problem} and \ref{sec:asymptotic_approx} to obtain the simple analytic forms:
\begin{equation}\label{eq:stability}
\eta=\left\{\begin{aligned}
        & 2\gamma\frac{\tau_g-\tau_l}{\tau_g+\tau_l}, &\quad \gamma < \omega \\
         & 2\gamma\left(\frac{\tau_g-\tau_l}{\tau_g+\tau_l}+ \sqrt{1-\left(\omega/\gamma\right)^2} \right),&\quad \gamma > \omega.
       \end{aligned}
 \right.
\end{equation}

\begin{figure}
\begin{center}
\includegraphics[scale=1]{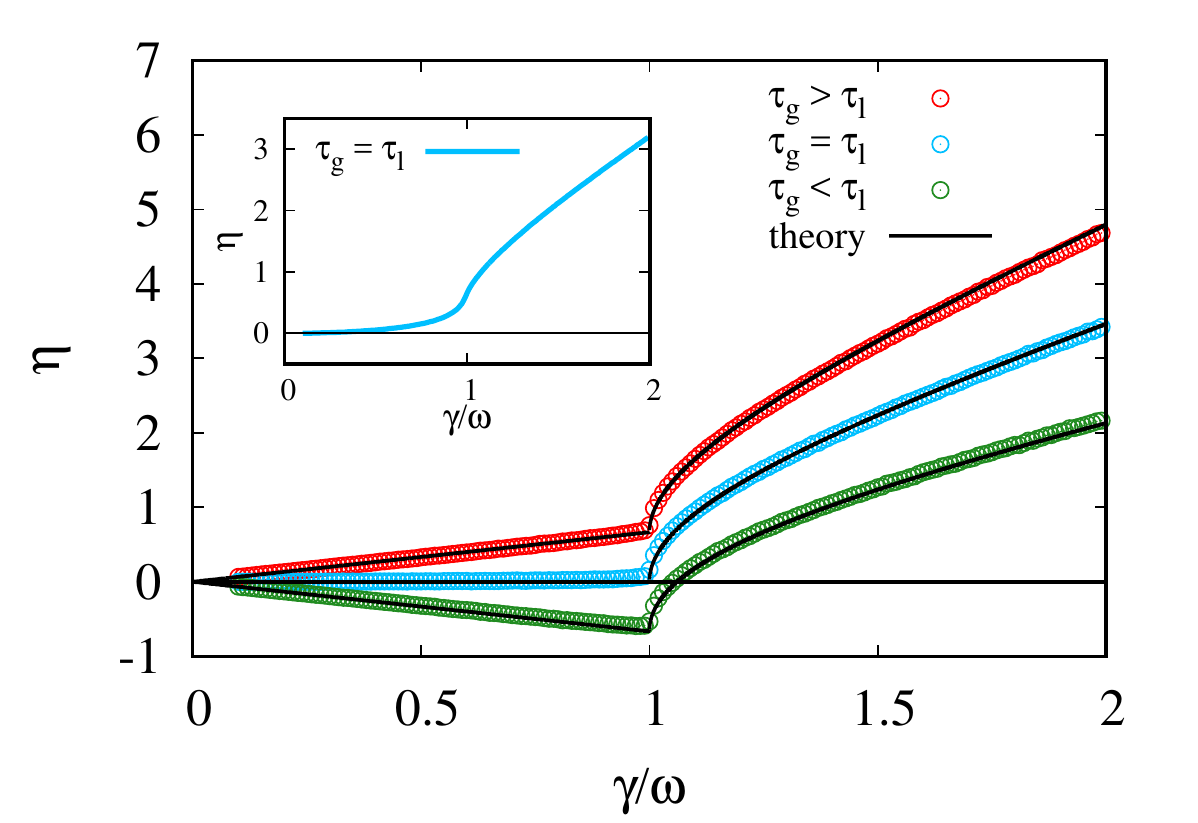} 
\end{center}
\caption{\textbf{Stability analysis of the stochastic oscillator as a function of $\bm{\gamma}$}. The colored circles correspond to values of the stability parameter $\eta$ in the large $\tau_l,\tau_g$ limit, determined numerically from simulations. \textbf{RED}: Predominant gain where $\tau_g=100$ and $\tau_l=50$. \textbf{BLUE}: Balanced state where $\tau_g = \tau_l = 50$. \textbf{GREEN}: Predominant loss where $\tau_g = 50$ and $\tau_l=100$. The \textbf{BLACK} curves correspond to the theoretical result displayed in (\ref{eq:stability}). The \textbf{INSET} shows the dependence of $\eta$ on $\gamma$ in the intermediate $\tau_g,\tau_l$ limit. The blue curve is the result of a simulation with $\tau_l=\tau_g=5$. In this limit, symmetry is broken even for $\gamma < \omega$. In all simulations an ensemble of $10^5$ oscillators evolving up to $t=300$  were used, each oscillator having the initial condition $(x_0,v_0)=(1,1)$ at $t=0$.}
\label{fig:stability}
\end{figure}

\noindent
From (\ref{eq:stability}), a necessary condition for stability is that $\tau_g < \tau_l$, independent of the values of $\omega$ and $\gamma$. Furthermore for $\gamma <\omega$, the asymptotic expression is in good agreement with the simulation results, whereby the oscillator is at the edge of stability for $\tau_g=\tau_l$. Asymptotically, 
the energy logarithm $\ln J$ can be considered as a deterministic quantity, which %[{\color{blue} \emph{Is it appropriate to use just energy or should one write %the logarithm of energy?}}] 
does not decay nor diverge when there is a perfect balance between gain and loss, when $\tau_g=\tau_l$. On the other hand, when $\gamma > \omega$ a gain-loss balance ($\tau_g=\tau_l$) does not induce stability in the system. On the contrary, the system 
%diver what is remarkable is the fact that the system 
can remain active with a growing or constant energy even when the loss states dominate over the gain states, i.e. when $\tau_l > \tau_g$. 
This is illustrated by the green curve in 
figure \ref{fig:stability}; there is a value of $\gamma$ above which the system's energy diverges no matter how large $\tau_l$ becomes compared to $\tau_g$.

\subsection{The effective transition to a broken $\mathcal{PT}$ 
symmetry phase}

We can more precisely formalize what was said above by investigating further the system by analyzing the symmetry properties of the probability density functions $p_g$ and $p_l$. 
We define the space reflection, or parity operator $\mathcal P$ as the exchange between the gain and loss probability densities of $\varphi$. In other words, $\mathcal P$ has the effect of swapping $g$ and $l$ so that we have the exchange $P_g\leftrightarrow P_l$.
A similar definition is encountered in \cite{bender2002generalized,bender2003must}
where for simplicity the real space is represented by a two-valued position 
(let us say $\pm 1$) and the parity operation represent the exchange 
between $+1$ and $-1$. This redefined operation has been used 
as a swap operation between two quantum states  and subsequently for 
the swap between the gain and loss states in \cite{tsironis2014ptZeroD}.  
We can also include the time-reversal operation $\mathcal T$, where 
$t \rightarrow -t$, $v\rightarrow -v$ and $x\rightarrow x$ so that $\varphi\rightarrow -\varphi$. If we apply both operations at the same time, the distributions $P_l(\varphi)$ and $P_g(\varphi)$ remain invariant so that the ${\cal PT}$ symmetry is fulfilled. 

\begin{figure}
\begin{center}
\includegraphics[scale=1]{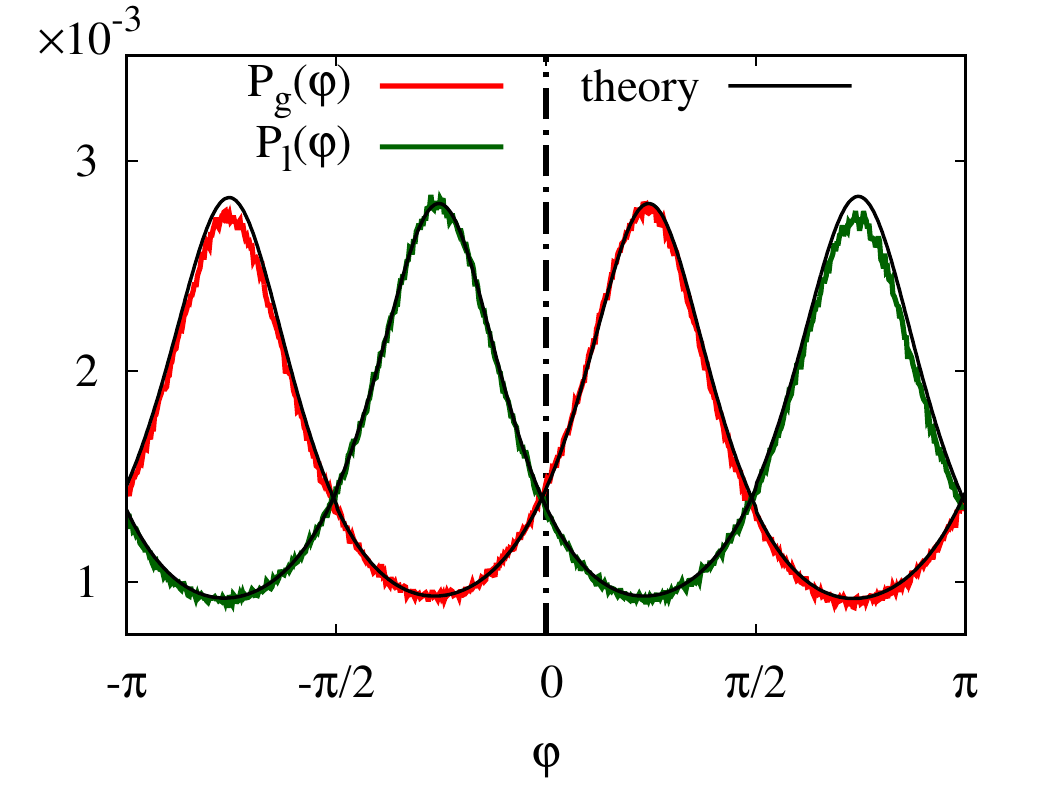}
\includegraphics[scale=1]{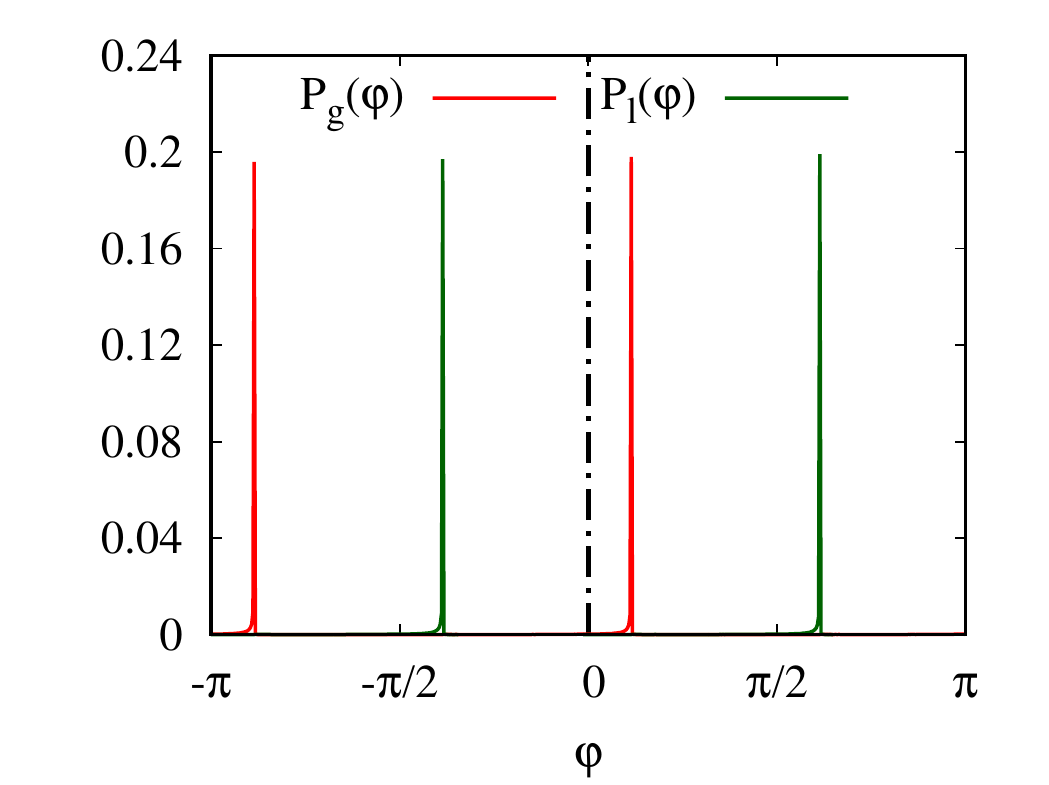}
\end{center}
\caption{\textbf{The distribution of the phases $\bm{P_g(\varphi)}$ and $\bm{P_l(\varphi)}$.} \textbf{TOP}: The system exhibits mirror symmetry and dispersed phases in the case where $\gamma < \omega$ and $\tau_g=\tau_L$.  In this example $\gamma=0.5$, $\omega = 1$ and $\tau_{g/l}=20$. An ensemble of $10^6$ oscillators was allowed to evolve up to $t=1500$. The black curve corresponds to the numerical solution of the analytical result in (\ref{eq:criterion2}).  \textbf{BOTTOM}: The symmetry is broken when $\gamma > \omega$ and the phases become localized. When $\gamma$ becomes larger than $\omega$ a transition occurs from a disordered to an ordered phase with a velocity-position ratio locking. In this example $\gamma=0.5$, $\omega = 1.5$ and $\tau_{g/l}=10$. An ensemble of $10^6$ oscillators was allowed to evolve up to $t=300$, each oscillator having the initial condition $(x_0,v_0)=(1,1)$ at $t=0$.}
\label{fig:symmetry}
\end{figure}

The ensemble of stochastic oscillators is effectively $\mathcal{PT}$-symmetric in the so-called {\it underdamped} regime when $\gamma<\omega$ and under the condition that $\tau_g=\tau_l$, although neither the stochastic equation (\ref{eq:damped_harm_osc}) nor the master equation  (\ref{eq:ME}) obey such a symmetry. Indeed, the phase probability densities $P_g(\varphi)$ and $P_l(\varphi)$ obtained by simulation and represented in the first graph of figure \ref{fig:symmetry} have a mirror symmetry with respect to the origin ($\varphi\rightarrow -\varphi$).
%with respect to the origin for $\tau_g=\tau_l$ %and under the additional condition that $\tau_g,\tau_l\gg 1$\todo{Is nt that valid for any value of equal tau's}. 
For comparison, the analytic expression for $P_{g/l}(\varphi)$ is obtained by solving the master equation in (\ref{eq:ME}) in the large time limit
(numerical integration and asymptotic expression in the large $\tau_{g/l}$ limit in \ref{sec:asymptotic_approx}).
%We determine the exact expression by numerically solving the integral in %(\ref{eq:X2}), which is the formal expression of the solution of %(\ref{eq:ME}) for large time.

However, this symmetry is only effective 
if we compare values of $\ln J$ up to 
its square root variance given that $\ln J$ keeps diffusing normally. 
But if we compare the square root variance relatively to any
non trivial average of $\ln J$, it shrinks to zero 
in the large time limit. 
Therefore these considerations have only a strict sense in the asymptotic limit viewed here as the analog of the thermodynamic limit where the concept of a large particle number of a thermodynamic system is replaced by one of large time, and where the so-called {\it normal} quantities are the average and the variance of $\ln J$ that both scale linearly with time (see \ref{sec:evolution_moments} and \ref{sec:criterion_appendix}). 

%\todo{Can you refer to your favorite book of thermodynamics where normal quantities concept is defined?}.   

The mirror symmetry of $P_{g/l}(\varphi)$ is maintained only for $\gamma < \omega$. Once this condition is no longer satisfied, the distribution of $\varphi$ initially broadly distributed in the symmetric case \emph{condenses} by forming two delta-like peaks.  It is in this sense that the system becomes deterministic once $\gamma$ is greater than $\omega$. At the same time, however, the mirror symmetry of the probability densities is broken as can be seen in the second graph of figure \ref{fig:symmetry} leading to a ${\cal PT}$-symmetry violation. In the limit of large $\tau_g$ and $\tau_l$ we are able to calculate a simple expression that determines the values $\varphi_{g/l}$ at which the two peaks in $P_{g/l}(\varphi)$ occur (see \ref{sec:asymptotic_approx} for details):
\begin{equation}\label{eq:ratio}
\tan\varphi_{g/l}=\omega\frac{x_{g/l}}{v_{g/l}}=\pm\frac{\gamma}{\omega}-\sqrt{\left(\frac{\gamma}{\omega}\right)^2-1}.
\end{equation}
The new "phase" obtained is ordered in the sense that it corresponds to a {\it ratio locking} of the velocity over the position with different fixed values for the gain state and the loss state. This result may also been obtained more intuitively by noticing that for  $\gamma \geq \omega$  the oscillator is damped with no oscillations. It corresponds to the {\it overdamped} regime as opposed to the underdamped regime where oscillations persist. We can indeed solve (\ref{eq:damped_harm_osc}) using the {\it ansatz} $x(t)=e^{\lambda t}x(0)$ and find that $\lambda=\pm \gamma \pm \sqrt{\gamma^2 -\omega^2}$ is real only in the overdamped regime. Hence, in the large time limit only one eigenvalue is dominant and therefore using $\omega x(t)/v(t)= \omega/\lambda$ for the dominant eigenvalue, we recover (\ref{eq:ratio}) accordingly. Such a relation could not have been used in the underdamped regime since the phase of oscillations 
would have randomized the trajectories. 

We interpret this observation as a phase transition from a {\it disordered state} to an {\it ordered state} with symmetry breaking in analogy to what happens in phase transition phenomena in thermodynamics. It can therefore be concluded that the ${\cal PT}$-symmetry breaking occurs at the point of critical damping ($\gamma=\omega$). In analogy to the Ising model \cite{toda1992statistical}, we start from a symmetric state with no ordering above a critical point, the broad angle distribution in our case (or the spin distribution in the magnet), and go towards a broken symmetry state with a well defined order with two possible opposite angle values e.g. ratio locking (or a well defined value of spin).

We end this section by adding that the symmetry breaking established in the limit of large $\tau_{l,g}$ is essentially valid also in the intermediate regime, despite a little bias ($\tau_l > \tau_g$) for $\gamma$ around $\omega$ shown in panel C of figure \ref{fig:stability2} and in the inset of figure \ref{fig:stability}. The symmetry is totally restored, however, in the limit of small $\tau_{l,g}$ whatever the value of $\gamma$ as can been seen from the inset in figure \ref{fig:stability2}.

%\begin{itemize}
%\item Relationship to Ising model
%\item How does the break in symmetry relate to the stability change of the system?
%\item The symmetry breaking occurs at the point of critical damping. When the system becomes overdamped its energy diverges when $\tau_g=\tau_l$.
%\end{itemize}

%\todo{I added a sentence in figure 6.} 

\section{Stability of the first moments}

\subsection{Dynamic equations for the position and velocity average} 

%As mentioned in the introduction, our main goal is to investigate the %stability of the stochastic oscillator  and this can also be done by
The stability criterion obtained in the previous section 
does not mean that all physical quantities of interest are stable. This statement can be illustrated by considering the evolution of the first moments of $p_g$ and $p_l$. The fact that $\ln J$ is stable does not necessarily mean that averages involving position and speed average are stable. Indeed, if $f(\varphi)$ is a function to average, from the Feymann-Gibbs inequality, we deduce:
\begin{equation}\label{fg}
\langle \alpha f(\varphi)\ln J\rangle \leq \ln \langle J^\alpha \exp(\alpha f(\varphi)
)\rangle
\end{equation}
On the contrary, the stability of position and velocity averages implies stability of $\ln J$. Therefore, there exist additional requirements that enhance the stability of the stochastic oscillators, adding to the richness of their dynamics.

By multiplying both equations in (\ref{eq:ME}) by $x$ and $v$ and integrating over the entire space and all the velocities, we obtain a system of coupled first order differential equations:
\begin{align}\label{eq:DS}
\frac{d}{dt}\langle x\rangle_l &= \langle v \rangle_l - \frac{\langle x \rangle_l}{\tau_l} + \frac{\langle x \rangle_g}{\tau_g}\\\notag
\frac{d}{dt}\langle v\rangle_l &= -\omega^2\langle x \rangle_l - 2\gamma\langle v \rangle_l - \frac{\langle v \rangle_l}{\tau_l} + \frac{\langle v \rangle_g}{\tau_g}\\\notag
\frac{d}{dt}\langle x\rangle_g &= \langle v \rangle_g + \frac{\langle x \rangle_l}{\tau_l} - \frac{\langle x \rangle_g}{\tau_g}\\\notag
\frac{d}{dt}\langle v\rangle_g &= -\omega^2\langle x \rangle_g + 2\gamma\langle v \rangle_g + \frac{\langle v \rangle_l}{\tau_l} - \frac{\langle v \rangle_g}{\tau_g}\notag
\end{align}
where 
\begin{equation}
\langle x \rangle_{g/l}=\int_{-\infty}^\infty dx\int_{-\infty}^\infty dv\; x\; p_{g/l}(x,v,t) \quad\text{and}\quad \langle v \rangle_{g/l}=\int_{-\infty}^\infty dx\int_{-\infty}^\infty dv\; v\; p_{g/l}(x,v,t).
\end{equation} 
The linear dynamical system described here is of the form $\bm{\dot z=M\cdot z}$ and therefore an asymptotically stable condition is verified when all of the real parts of the roots of the characteristic polynomial associated with $\bm M$ are negative. It is straightforward to determine the four eigenvalues which we shall denote as $\bm\lambda = (\lambda_1,\lambda_2,\lambda_3,\lambda_4)$ (see \ref{sec:first_moment_evalue} for details). Consequently the stability is also asymptotic in time. 

In a way similar to what was presented in the previous sections, the dynamical system governing the first moments can be separated into two regimes, namely, the case where $\gamma < \omega$ and $\gamma > \omega$. We proceed by examining the stability conditions for such a system and without loss of generality we assume that $\omega=1$. We show that in this case also, depending on the value of $\gamma$, it is possible to have a situation in which the average quantities we studied diverge even though the system, on average, spends more time in the loss state than in the gain state.
In contrast to the results obtained regarding the stability of $\ln J$ in the previous section, the average values in (\ref{eq:DS}) diverge when $\tau_l=\tau_g$ since there exists at least one eigenvalue 
that is positive for any combination of the other parameters 
%$\gamma$ and $\tau_g=\tau_l$ 
(see \ref{sec:first_moment_evalue} for details).

%We show that in this case also, depending on the value of $\gamma$, it is possible to have a situation in which the dynamics of the oscillator diverges even though, over time, on average it loses more energy than it receives. \todo{This last sentence is not necessarily true if one does not study the stability of $<J>$}

\subsection{Underdamped case: $\gamma < \omega$}

%In the underdamped case, where $\gamma < \omega$
In this case, all four eigenvalues $\lambda_i$ are complex and they come in conjugate pairs (see \ref{sec:first_moment_evalue}). The real parts of two of them are equal, $Re\lambda_1=Re\lambda_2$, and negative for all combinations of 
$(\tau_g$,$\tau_l)$
%\in \mathbb R^+\times\mathbb R^+$ 
whilst the real parts of the other two, which are also equal ($Re\lambda_3=Re\lambda_4$), can be either positive or negative. 
Since the imaginary parts of all four eigenvalues are always non-zero, 
all the solutions of the system (\ref{eq:DS}) are oscillatory, whether they decay or grow, for any value of $\tau_l$ and $\tau_g$.
For a fixed $\gamma$ we numerically determine the regions in the $\tau_l-\tau_g$ parameter space for which every eigenvalue of the system has a negative real part. This corresponds to the stable region and is indicated in green in panels A and B of figure \ref{fig:stability_average}.
%and \ref{fig:stability_average}, 
%where we see that stability is possible only in cases where $\tau_l>\tau_g$.
Moreover, the relation
$\lim_{\tau_l\rightarrow\infty}\lambda_i = \gamma -1/\tau_g$ for $i=3,4$ 
derived using 
\ref{eq:eigenvalues} 
imposes the critical value $\tau_g^*= 1/\gamma$ beyond which stability is never attained no matter how large the value of $\tau_l$ is.

%for a fixed value of $\gamma$, stability is never attained when %$\tau_g>1/\gamma$ no matter how large the value of $\tau_l$ is. 
%Indeed, the limit relation
%$\lim_{\tau_l\rightarrow\infty}\lambda_i = \gamma -1/\tau_g$ for $i=3,4$  %imposes a bound at $\gamma=1/\tau_g$.

%Therefore, the eigenvalues $\lambda_3$ and $\lambda_4$ change the stability of %(\ref{eq:DS}) by switch sign at $\gamma=1/\tau_g$. 

%\todo{too much redundancy, shorten the two paragraphs below and go straight to the point}
%Indeed, when $\gamma < \omega$ it is enough to investigate the real part of $\lambda_3$ (or $\lambda_4$) in order to determine the maximum value of $\tau_g$ for which there exists a stable solution. For $\tau_l$ that tends to infinity, we find that (see \ref{sec:first_moment_evalue}) $\lim_{\tau_l\rightarrow\infty}\lambda_3 = \gamma -1/\tau_g$, which proves that for $\gamma<\omega$, stability is never attained when $\tau_g>1/\gamma$, independently of $\omega$. In addition, we can deduce that the boundary separating the stable from the unstable region shifts to smaller values of $\tau_g$ as $\gamma$ increases.

%Therefore the system can never be stable in the symmetric case.

\subsection{Overdamped case:  $\gamma \geq \omega$}

\begin{figure}[htp]
\begin{center}
\includegraphics[width=1\linewidth]{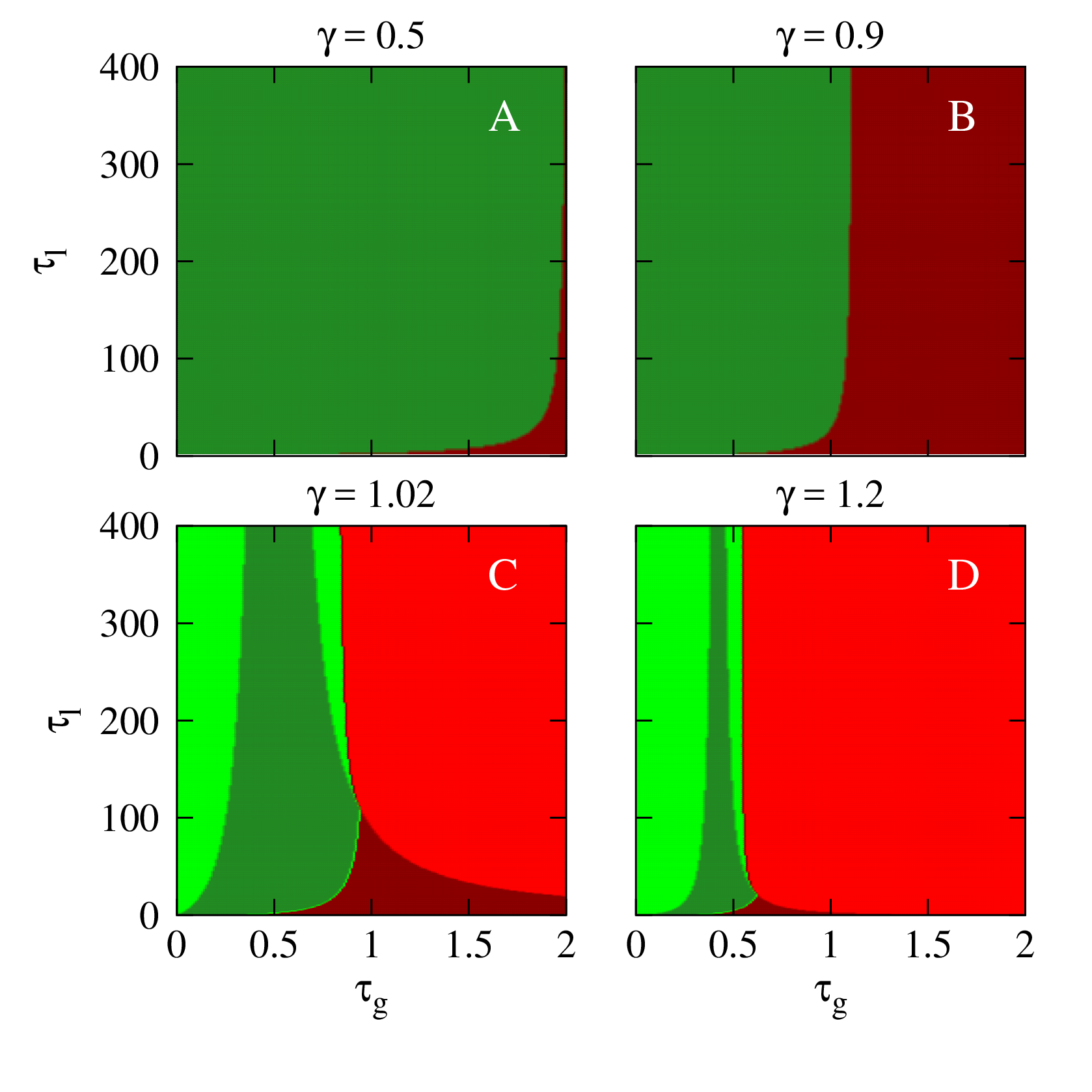} 
\end{center}
\caption{\textbf{Stability analysis for the evolution of the first moments.} The stable regions in parameter space for which all four eigenvalues have negative real parts are colored green while the unstable regions are in red. Each panel in the figure was made by evaluating directly the real parts of the eigenvalues in (\ref{eq:eigenvalues}) belonging to the dynamical system (\ref{eq:DS}). In all cases $\omega=1$ and each oscillator was with the initial condition $(x_0,v_0)=(1,1)$ at $t=0$. In the underdamped regime, the comparison between \textbf{ PANELS A} and \textbf{B} show a stable region that decreases with growing $\gamma$. \textbf{PANELS C} and \textbf{D} show two examples of the overdamped regime for different values of $\gamma$. The dark green and dark red colors 
%refers for stable and unstable regions  
denote the set of values $(\tau_g,\tau_l)$ for which the solutions of the dynamical system are oscillatory (in the underdamped regime, all solutions are oscillatory). In this case also, both the stable and oscillatory regions diminish in size with increasing $\gamma$.}
\label{fig:stability_average}
\end{figure}

Once the stochastic oscillator is overdamped, more of the eigenvalues can attain positive real parts, rendering the dynamics more elaborate.
%richer and more complex. 
In particular, in contrast to the underdamped case, $Re\lambda_2$ can also be positive, depending on the values of $\tau_g$ and $\tau_l$. Nevertheless, the stability does not differ qualitatively from the underdamped case. The boundary that separates the stable from the unstable region continues to shift towards lower values of $\tau_g$ when $\gamma$ increases further. %{\color{blue}In addition, for reasons already mentioned for the case $\gamma %< \omega$, stability is never attained in the symmetric case where %$\tau_l=\tau_g$.}
%The particular condition where $\tau_l=\tau_g$ is characterized by (\ref{eq:lambda_symmetric}) also in the overdamped case. Therefore, we can conclude that stability is never attained in this case either when $\tau_l=\tau_g$. 
%\todo{Finish this paragraph. \color{red} Ok finish it and I will go through this after. Try to avoid redun{dancy or repetition.}}
Another interesting property deduced from the eigenvalues is that in the overdamped case there exist oscillatory solutions to (\ref{eq:DS}) 
in contrast to the simple damped harmonic oscillator that is  
monotonically damped  under those conditions. These oscillations occur when at least one of the eigenvalues has a non-zero imaginary part. 
%When $\gamma > \omega$, 
This is the case for values of $(\tau_l,\tau_g)$ for which $|Im\lambda_1|+|Im\lambda_2|+|Im\lambda_3|+|Im\lambda_4|\neq0$. 
%A solution to this problem was determined numerically and is presented in figure \ref{fig:oscillation}. 
These results are presented in panels C and D of figure \ref{fig:stability_average}
where, although the monotonic  motion is predominant in the overdamped case, there persists a region in the $\tau_g-\tau_l$ plane for which the solutions are oscillatory. 

\section{Full time analysis at zero frequency}

Until now, we have analyzed the asymptotic behavior of the oscillator in the large time limit and also the first moment of the position and the velocity. For the sake of completeness, it remains in principle to discuss about the short and intermediate time regimes and the stability of the higher order moments.
A full analysis  is beyond the scope of this paper but 
we can address the particular case of $\omega=0$, so that the problem 
reduces to the simpler dynamical equation in which the position coordinate is 
eliminated:
\begin{eqnarray}\label{vel}
\frac{d v(t)}{dt}=\pm 2\gamma v(t)
\end{eqnarray}
This dynamical equation describes the evolution of a wave function 
that is amplified or undergoes loss randomly. A similar study where the frequency takes randomly two possible values has been shown useful in the context 
of superconducting qubits \cite{Paraoanu2013}. 
This type of model has already been used to determine the first 
passage time at which, for instance, the speed exceeds a critical value 
\cite{Tsironis1988,Masolivera1986,Masoliverb1986,Masoliver1987}.
The interest here is to illustrate how the exact solution complements the results obtained so far. 

We start from the initial condition: $p_{g/l}(v,t=0)=p_{g/l}\delta(v-v_0)$
and solve this equation using the variable change: $r=\ln(v/v_0)/(2\gamma)$
(see \ref{omega0}).
The short time analysis  differs from the asymptotic analysis because the distribution is not normal anymore. If we start with a well defined speed 
$v_0$ and study its subsequent evolution for short time ($t \ll \tau_{g/l}$), we obtain the spread of the velocity distribution but  
confined within a cone:
\begin{eqnarray}\label{shorttime}
p_{g/l}(v,t\simeq 0)=\frac{1}{v}\left[p_{g/l}\delta(r-t)(1-\frac{t}{\tau_{g/l}})
+\frac{p_{l/g}}{2\tau_{l/g}}1^+(t^2-r^2)\right],
\end{eqnarray}
where $1^+(x)$ is the Heaviside function. 
The short time behavior is characterized by a 
propagation of a delta distribution along this cone inside 
which the probability densities develop. 
In the opposite case for large time ($t \gg \tau_{g/l}$), we find a normal distribution for 
the variable $r$ with  average and variance:
\begin{eqnarray}\label{av}
\langle \ln\left(\frac{v(t)}{v_0}\right) \rangle=
\frac{\tau_g-\tau_l}{\tau_g+\tau_l}2\gamma t,
\quad
\sigma^2(t)=
\langle \delta^2 \ln\left(\frac{v(t)}{v_0}\right)\rangle= \frac{(2\gamma)^28\tau_l^2 \tau_g^2}{(\tau_g +\tau_l)^{3}}t
\end{eqnarray}
Thus, the stability of the average is satisfied for $\tau_l \geq \tau_g$ in contrast to the previous statements in (\ref{eq:stability}), which always predict instability. 
The apparent contradiction is resolved by remembering that according to 
(\ref{fg}) the stability of $r$ does not imply the stability of any 
moment of the velocity and/or any moment of the position. Indeed 
 for the $n^{th}$ moment, 
we determine the following more restrictive stability criterion:
\begin{eqnarray}\label{n}
n \leq \frac{\tau_l-\tau_g}{2\gamma\tau_g\tau_l}
%\geq 0  \quad \quad  \alpha \in R
\end{eqnarray}
Therefore, there always exists an order $n$ above which the stability criterion 
is not satisfied in accordance with the variance in (\ref{av}), which always
increases. 
However, if we restrict to the $v$ variable only without the position $x$ 
then the simplified system becomes effectively always ${\cal PT}$ symmetric 
when $\tau_l=\tau_g$.

%\begin{eqnarcomesray}\label{moment}
%\ln\frac{\langle v^n(t)\rangle}{v_0^n} 
%\stackrel{t \rightarrow \infty}{=}
%\left(\sqrt{\left(\frac{1}{\tau_g}+\frac{1}{\tau_l}\right)^2
%+16(\gamma n)^2-8\gamma n\left(\frac{1}{\tau_g}-\frac{1}{\tau_l}\right)}
%-\frac{1}{\tau_g}-\frac{1}{\tau_l}\right)\frac{t}{2} 
%\quad
%\end{eqnarray}

\section{Conclusions and perspectives}

We have studied the dynamic evolution of stochastic oscillators subject to dichotomous noise made of alternating gain and loss states random in time and we have unveiled an intimate connection of this non conservative system with ${\cal PT}$ symmetry. We established a useful criterion that fixes the boundary line between  a stable regime with a likely decaying amplitude and an unstable regime with a likely growing one.
Although the oscillator evolution becomes more stochastic with time, 
it is nevertheless possible to effectively define the useful concept of ${\cal PT}$ symmetry in the asymptotic time limit. 
In other words, despite the breaking of time reversal invariance due to noise, the oscillator can still remain  resilient so as to preserve at least the ${\cal PT}$ symmetry. 
Application of this invariance property allows to distinguish between different regimes or phases: a) an underdamped regime (or weakly damping-amplifying oscillator ) for which the boundary lines between stable and unstable regions satisfy this symmetry; b) an overdamped regime (or strongly damping-amplifying oscillator) for which this boundary line becomes asymmetric. We interpret these results in analogy to thermodynamics as a phase transition from a symmetric disordered state consisting of a broad distribution to an ordered state with a restricted distribution imposing a {\it ratio locking} of the position over the velocity separately for both the gain and loss states.

To complete the panorama, we also examined the time evolution of the position and velocity averages of the oscillator. We showed that the stability of the oscillator does not necessarily imply bounded dynamics of these averages. It appears indeed that the stability diagrams are more elaborate, illustrating the much richer structure of this apparently very simple system. For instance, despite the absence of oscillations in the deterministic case, in the overdamped regime the presence of a random gain can re-stimulate them. Higher order moment analysis together with a study of the short and intermediate time limits confirm this broad range of different regimes with different physics such as the cone-like propagation of the velocity distribution.

The formalism developed here in the particular case of a stochastic oscillator is quite general and may be applied to other situations where 
dichotomous noise is present such as the stabilization of
light propagation in metamaterials and optical fibres with random regions of
asymmetric active and passive media \cite{ruter2010observation}.

%{\color{blue} TO DO
%\begin{itemize}
%\item We have shown the stability criterion for an ensemble of infinite oscillators. Perhaps, it could be possible to develop a model for a finite number of oscillators. This might be closer to real systems than the ideal situation considered in the manuscript.

%{\color{red} this becomes a stastitic problem. Idon't see the interest of 
%doing. it looks like playing tie with finite shots}

%\item Intuitive explanation of the instability of the system even when %$\tau_l > \tau_g$ for $\gamma > \omega$. Discuss the fact that we are dealing %with average values over an ensemble of stochastic oscillators and that a %single one diverges, it pulls the average towards higher values.

%{\color{red} I have the answers}

%\item Effects of having a finite number of oscillators.

%{\color{red} 
%There is not effect since oscillator do notinteract }

%\end{itemize}
%}

\ack
We acknowledge partial supports of the European Union's Seventh Framework Programme (FP7-REGPOT-2012-2013-1) under grant agreement number 316165 and by the EPSRC grant EP/M006581/1. M.L. acknowledges financial support from the ERC Advanced grant number FP7-319968 FlowCCS of the European Research Council. Helpful discussions with K. Makris is also gratefully acknowledged.
%\todo{Karl Bender also unless we want him as a referee} 

%and by the Ministry of Education and Science of the Russian Federation in the framework of Increase Competitiveness Program of NUST MISiS No. K2-2015-007.

\newpage
\appendix

\section{Asymptotic solution of the master equation}

\subsection{Dominant eigenvalue}\label{sec:dominant_eigenvalue}

After reducing the master equation (\ref{eq:ME}) to a pair of coupled ordinary differential equations by means of integral transforms, 
we write the master equation (\ref{eq:ME_transform}) under a matrix form
with $p_{g/l}(J,\varphi,t_0)$ as the initial distribution:
\begin{equation}
\mathcal M\cdot \hat p=-s\hat p + \hat p(k,\varphi,t_0),
\end{equation}
where 
\begin{equation}\label{eq:eigenvalue_problem}
\mathcal M = \mathcal M^{(0)}+k\mathcal M^{(1)}=
\left[\begin{pmatrix} \mathcal{L}_g^{(0)} & -1/\tau_l \\ -1/\tau_g & \mathcal{L}_l^{(0)} \end{pmatrix}  + k\begin{pmatrix} -4\gamma\cos^2\varphi & 0 \\ 0 & 4\gamma\cos^2\varphi \end{pmatrix}\right] 
\end{equation}
in which
\begin{equation}
\mathcal{L}^{(0)}_{g/l} = \frac{d}{d\varphi}(\omega \mp\gamma\sin2\varphi) + \frac{1}{\tau_{g/l}}.
\end{equation}
and where 
\begin{equation}
\hat p = 
\left( \begin{array}{c} \hat p_{g} \\ \hat p_{l} \end{array} \right)
\quad \quad
\hat p(k,\varphi,t_0)=
\left( \begin{array}{c} \hat p_{g}(k,\varphi,t_0) \\ 
\hat p_{l} (k,\varphi,t_0)\end{array} \right).
\end{equation}
Note that because of the inversion symmetry $x \rightarrow -x$ and 
$v \rightarrow -v$, the stochastic oscillator is invariant  
under the transformation $\varphi \rightarrow \varphi +\pi$. Therefore, we can restrict the angle to the interval $\varphi \in ]-\pi/2,\pi/2]$.
Writing the solution $\hat p(k,\varphi,s)$ as a linear combination of the eigenfunctions $\hat p_i(k,\varphi)$ of $\mathcal M$ with eigenvalue $-s_i$ we obtain
\begin{equation}\label{eq:hatP}
\hat p(k,\varphi,s) = \sum_i c_i(s) \hat p_i(k,\varphi).
\end{equation}
In that case, the master equation becomes
\begin{equation}
\sum_i (s+s_i)c_i(s) \hat p_i(k,\varphi) = \hat p(k,\varphi,t_0).
\end{equation}
Multiplying from the left  with the eigenfunction $\tilde p_j^\dagger(k,\varphi)$ 
of the adjoint problem with eigenvalue $s_j$ and using 
the bi-orthogonality property
\begin{equation}\label{eq:biortho}
\int_{-\pi/2}^{\pi/2} \tilde p_i^{\dagger}\cdot \hat p_j^{}\;d\varphi = 0 
\quad \quad i\not=j,
\end{equation}
we obtain
\begin{equation}
c_j(s)=\frac{\int_{-\pi/2}^{\pi/2} \tilde p_j^\dagger(k,\varphi) \cdot \hat p(k,\varphi,t_0)\; d\varphi
}{(s+s_j) \int_{-\pi/2}^{\pi/2} \tilde p_j^\dagger(k,\varphi)\cdot \hat p_j(k,\varphi)\; d\varphi}=\frac{A_j(k)}{s+s_j}.
\end{equation}
Taking the inverse Laplace transform, we arrive at
\begin{equation}
c_j(t) = \exp\left[-s_jt\right] A_j(k).
\end{equation}
Therefore, the inverse Laplace transform of (\ref{eq:hatP}) gives
\begin{equation}
\hat p(k,\varphi,t) = \sum_i \exp[-s_jt]A_j(k)\hat p_i(k,\varphi).
\end{equation}
In the asymptotic time limit where $t\rightarrow\infty$, assuming that $s_0$ is the dominant eigenvalue in the sense that the real component 
${\rm Re}\,s_0$ has the lowest value among all eigenvalues,
%[{\color{blue} \emph{It has to be justified why $s_0$ is the largest eigenvalue in %the sense that $-{\rm Re}s_0$ is the largest value. . Perhaps because $s$ has to be %positive since $t$ has to be positive and hence $-s \leq 0$? Also the fact that %$s=0$ corresponds to the asymptotic time limit.}}], 
we simplify the dynamics into:
\begin{equation}\label{eq:Pasymptotic}
\hat p(k,\varphi,t\rightarrow\infty) = \exp[-s_0t]A_0(k)\hat p_0(k,\varphi).
\end{equation}

\subsection{Perturbation theory}\label{sec:perturbation_theory}

%After reducing the master equation in (\ref{eq:ME}) to a pair of coupled ordinary %differential equations by means of integral transforms, 
%\begin{equation}\label{eq:hatP}
%\hat p_{g/l}(k,\varphi,s) = \sum_i c_i(s) \hat p_{i,g/l}(k,\varphi)
%\end{equation}
%where $\hat p_i(k,\varphi)$ is the solution of the eigenvalue 
%problem:
%we still have no other option but to resort to perturbative methods. We therefore %rewrite the master equation int  eigenvalue problem in matrix form: [{\color{blue} %\emph{The starting distribution is missing. It should be either incorporated or %removed with a justification.}}]
%\begin{equation}\label{eq:eigenvalue_problem}
%\left[\begin{pmatrix} \mathcal{L}_g^{(0)} & -1/\tau_l \\ -1/\tau_g & %\mathcal{L}_l^{(0)} \end{pmatrix}  + k\begin{pmatrix} -4\gamma\cos^2\varphi & 0 \\ %0 & 4\gamma\cos^2\varphi \end{pmatrix}\right] \cdot 
%left( \begin{array}{c} \hat p_{i,g} \\ \hat p_{i,l} \end{array} \right) = -s_i
%\\left( \begin{array}{c} \hat p_{i,g} \\ \hat p_{i,l} \end{array} \right)
%\end{equation}
%where
%\begin{equation}
%\mathcal{L}^{(0)}_{g/l} = \frac{d}{d\varphi}(\omega \mp\gamma\sin2\varphi) + %\frac{1}{\tau_{g/l}}.
%\end{equation}
%with eigenvalue $s_i$ labeled by $i$.
%Consequently, we can treat (\ref{eq:eigenvalue_problem}) as an eigenvalue %perturbation problem of the from 

We progress further using the perturbation expansion applied to the eigenvalue problem:
\begin{equation}\label{eq:ME_operator_form}
\mathcal M \cdot \hat p_i = (\mathcal M^{(0)}+k\mathcal M^{(1)})\cdot\hat p_i=-s_i\hat p_i
\end{equation} 
where $\hat p_i = (\hat p_{i,g} \;\hat p_{i,l})^T$ is a spinor and where the matrix operator $\mathcal M$ is the sum of the principle part $\mathcal M^{(0)}$ and a perturbation part $\mathcal M^{(1)}$ whose matrix elements can be deduced directly from (\ref{eq:eigenvalue_problem}).
In order to solve the eigenvalue problem for the operator (\ref{eq:eigenvalue_problem}), we expand the $i^{th}$ eigenvalue, $s_i$, of the operator $\mathcal M$ in powers of the perturbation parameter $k$ so that 
\begin{equation}
s_i=s_i^{(0)}+ks_i^{(1)}+k^2s_i^{(2)}/2!+\ldots\;.
\end{equation}
In the leading order,  $s_i^{(0)}$ is equal to the eigenvalue of the unperturbed operator $\mathcal M^{(0)}$ while, in the first order, we find using (\ref{eq:biortho}) the first correction:
%and that [{\color{blue}\emph{explain why the limit of integration is $\pi/2$, i.e. %why the period is $\pi/2$.}}]
\begin{equation}\label{eq:firstorder}
s_i^{(1)}=-\frac{\int_{-\pi/2}^{\pi/2}\tilde p_i^{\dagger(0)}\cdot \mathcal M^{(1)}\cdot \hat p_i^{(0)}d\varphi}{\int_{-\pi/2}^{\pi/2}\tilde p_i^{\dagger(0)}\cdot \hat p_i^{(0)}d\varphi},
\end{equation}
where $\hat p_i^{(0)}=(\hat p_{i,g}^{(0)} \; \hat p_{i,l}^{(0)})^T$ 
%=(p_{i,g}(0,\varphi)\; \hat p_{i,l}(0,\varphi))^T$ 
is the solution to the unperturbed eigenvalue problem ($i^{(th)}$ eigenvector) that involves only $\mathcal M^{(0)}$. Since the operator is not self-adjoint, we must use $\tilde p_i^{(0)}$, which is the solution to the zeroth order adjoint problem defined by
\begin{equation}
\mathcal M^{\dagger(0)}\cdot\tilde p^{(0)}_i = \begin{pmatrix} \mathcal L^{\dagger(0)}_g & -1/\tau_g \\ -1/\tau_l & \mathcal L^{\dagger(0)}_l \end{pmatrix} \cdot \left( \begin{array}{c} \tilde p_{i,g}^{(0)} \\ \tilde p_{i,l}^{(0)} \end{array} \right) = -s^{(0)}_i\left( \begin{array}{c} \tilde p_{i,g}^{(0)} \\ \tilde p_{i,l}^{(0)} \end{array} \right),
\end{equation}
where the adjoint operator has the form
\begin{equation}
\mathcal{L}_{g/l}^{\dagger(0)} = -(\omega \mp\gamma\sin2\varphi)\frac{d}{d\varphi} + \frac{1}{\tau_{g/l}}.
\end{equation}
It is easily verified that in the unperturbed case, the adjoint problem 
has the trivial solution $\tilde p_0^{(0)}=(1 \; 1)^T$ with the eigenvalue  $s_0^{(0)}=0$. This trivial eigenvalue 
is the most dominant because an eigenvalue with a real positive value would lead
to a probability that is not conserved in time. In fact, this eigenvalue 
is precisely associated to the probability conservation.
%In that case \ref{eq:biortho} we have that
%\begin{align}
%s_0^{(1)}=\mathcal M_{00}^{(1)}&\equiv\int_{-\pi/2}^{\pi/2}\tilde p_0^{\dagger(0)}\mathcal M^{(1)}\,\hat p_0^{(0)}d\varphi\notag\\
%&=\int_{-\pi/2}^{\pi/2}-4\gamma\cos^2\varphi\;\left(\hat p_g(0,\varphi,0)-\hat p_l(0,\varphi,0)\right)d\varphi.
%\end{align}
Therefore, up to first order in $k$, the dominant solution 
of the eigenvalue problem (\ref{eq:firstorder}) is rewritten using (\ref{eq:eigenvalue_problem}) more simply as:
\begin{equation}\label{eq:s_0}
s_0 = s_0^{(0)}+ks_0^{(1)}=-k\frac{\int_{-\pi/2}^{\pi/2}4\gamma\cos^2\varphi\;\left(\hat p_g(0,\varphi,0)-\hat p_l(0,\varphi,0)\right)d\varphi}{\int_{-\pi/2}^{\pi/2}\left(\hat p_g(0,\varphi,0)+\hat p_l(0,\varphi,0)\right)d\varphi}.
\end{equation}

%It is possible, in principle, to formally write down the solution $\hat p^{(0)}(0,\varphi,0)$ of the unperturbed case ($k=0$) of (\ref{eq:eigenvalue_problem}) for $s=0$ in terms of the corresponding integrating factor. However, working through all the necessary integrals can be extremely difficult, especially in the case $\gamma > \omega$ for which the first order coupled differential equation becomes singular. {\color{blue} \emph{There should be a sentence here stating that $\hat p(k,\varphi,s)$ can be written as a superposition of all the unperturbed solutions, including $s>0$ but that it is simply too difficult and not worth doing, since we are interested in the asymptotic case.}} Instead of searching for the approximate expression for the solution of (\ref{eq:eigenvalue_problem}) in terms of the unperturbed eigenvectors, we focus on the asymptotic stability of the solution. It is straightforward to show via (see \ref{sec:evolution_moments}) the moment generating function that the $n^{th}$ central moment of the quantity $\ln J$ has the following expression in the asymptotic time limit

\subsection{Evolution of the central moments of $\ln J$}\label{sec:evolution_moments}

We start with the Mellin transform shown in (\ref{eq:transforms}) and write it down as a characteristic function of the total probability 
%$p_g(J,\varphi,t) + p_l(J,\varphi,t)$ 
for the variable $\ln J$:
\begin{eqnarray}
M_{\ln J}(k,t)&=&
\int_{-\pi/2}^{\pi/2} \!\!
\hat p_g(k,\varphi,t) + \hat p_l(k,\varphi,t)\;d\varphi
\nonumber \\
&=&
\int_{-\pi/2}^{\pi/2} \int_0^\infty \!\! e^{k\ln J} (p_g(J,\varphi,t)+ p_l(J,\varphi,t))\;dJ d\varphi .
\end{eqnarray}
If we write down the exponential as a power series and integrate over $\varphi$, we obtain an expression with the moment generating function of $\ln J$ on the right-hand-side:
\begin{equation}
%\int_{-\pi/2}^{\pi/2}\hat p_g(k,\varphi,t) + \hat p_l(k,\varphi,t)\;d\varphi = 
M_{\ln J}(k,t)=
\sum_{n=0}^\infty\frac{k^n}{n!}\langle(\ln J)^n\rangle(t).
\end{equation}
Therefore, in order to obtain the evolution in time of the $n^{\rm th}$ \textit{central} moment of $\ln J$, we take the $n^{\rm th}$ derivative of  the logarithm of the moment-generating function above (i.e. derive the cumulant generating function) with respect to $k$ \cite{vanKampen2007stochastic},
\begin{equation}
%\mu_n(t) = 
\langle(\ln J -\langle\ln J\rangle)^n\rangle = \left.\frac{d^n}{dk^n}\right\vert_{k=0}\ln M_{\ln J}(k,t).
%\left[\int_{-\pi/2}^{\pi/2}\hat p_g(k,\varphi,t) + \hat %p_l(k,\varphi,t)\;d\varphi\right].
\end{equation}
According to perturbation theory, we can expand the dominant eigenvalue $s_0$ in powers of $k$. By substituting this expansion into the Mellin-transformed probability distribution obtained in the asymptotic time limit in equation (\ref{eq:Pasymptotic}) we obtain:
\begin{eqnarray}
\ln M_{\ln J}(k,t)&\stackrel{t\rightarrow \infty}{=}&
-s_0^{(0)}t-ks_0^{(1)}t-k^2s_0^{(2)}t/2! -\ldots + \ln A_0(k)
\nonumber \\
&+&
\ln  \left(\int_{-\pi/2}^{\pi/2}
\hat p_{0,g}(k,\varphi) + \hat p_{0,l}(k,\varphi)\;d\varphi\right) .
\end{eqnarray}
Finally, for the $n^{th}$ central moment with respect to the quantity $\ln J$, we identify simply in the asymptotic time limit:
\begin{equation}\label{eq:nth_moment}
%\mu_n(t)
\langle(\ln J -\langle\ln J\rangle)^n\rangle \stackrel{t\rightarrow \infty}{=}
-s_0^{(n)}t.
\end{equation}
where $s_0^{(n)}$ is the $n^{th}$ expansion coefficient in $k$ 
of the eigenvalue $s_0=\sum_{n=0}^\infty {s_0^{(n)}k^n}/{n!}$.
Note that the asymptotic result does not depend anymore on the initial 
condition through $A_0(k)$.

\subsection{Stability criterion}\label{sec:criterion_appendix}

As already mentioned in \ref{sec:perturbation_theory}, the master equation can be treated as a perturbed eigenvalue problem with $k$ as the perturbation parameter and $s$ as the eigenvalue to be determined. 
%It is straightforward to show via the moment generating function (see \ref{sec:evolution_moments}) that the $n^{th}$ central moment of the quantity $\ln J$ has, in the asymptotic limit,\stackrel{t\rightarrow \infty}{=}
%\begin{equation}\label{eq:nth_moment}
%\mu_n(t) = \langle(\ln J -\langle\ln J\rangle)^n\rangle = -s_0^{(n)}t,
%\end{equation}
%where $s_0^{(n)}$ is the $n^{th}$ order approximation of the eigenvalue $s_0$ [{\color{blue} \emph{Check whether this last phrase is written correctly}}]. 
Given that the dominant eigenvalue for the unperturbed problem is $s^{(0)}_0 = 0$ (see \ref{sec:dominant_eigenvalue}), so that $s_0=ks_0^{(1)}$ up to first order, we can substitute the time derivative of the first moment in (\ref{eq:nth_moment}) into (\ref{eq:s_0}) to obtain finally the stability criterion defined in section \ref{sec:stability_criterion}:
\begin{equation}\label{eq:criterion_fourier}
\frac{d}{dt}\langle \ln J \rangle(t) \stackrel{t\rightarrow \infty}{=}
-s_0^{(1)} = \frac{\int_{-\pi/2}^{\pi/2}4\gamma\cos^2\varphi\;\left(\hat p_g(0,\varphi,0)-\hat p_l(0,\varphi,0)\right)d\varphi}{\int_{-\pi/2}^{\pi/2}\left(\hat p_g(0,\varphi,0)+\hat p_l(0,\varphi,0)\right)d\varphi}.
\end{equation}
The linear growth of the first two central moments means that the \emph{relative square root variance} 
or the square root variance-to-mean ratio 
shrinks to zero asymptotically, i.e.,
\begin{equation}
\frac{\sqrt{Var[\ln J]}}{\langle \ln J\rangle}
%=\frac{\mu_2}{\mu_1}
\stackrel{t\rightarrow \infty}{=}\frac{\sqrt{s_0^{(2)}}}{s_0^{(1)}}
\frac{1}{\sqrt{t}} \rightarrow 0,
\end{equation}
provided $\langle \ln J\rangle \not= 0$.
This result leads us to conclude that $\ln J$ is in general a well defined statistical variable and can be viewed as a deterministic one in the asymptotic sense (see figure \ref{fig:variance}). Strictly speaking,
only the case $\langle \ln J \rangle = 0$ has to be considered non deterministic but distributed within the interval of the square root variance.

\subsection{Solution to the unperturbed eigenvalue problem}\label{sec:unperturbed_problem}

Setting $s=k=0$ in (\ref{eq:ME_transform}), 
we notice that the functions:
%and adding and subtracting the two equations in (\ref{eq:ME_transform}) we obtain
%\begin{equation}
%\frac{d}{d\varphi}X_1 = 0 \quad\text{and}\quad \frac{d}{d\varphi}X_2 = %-\left[\frac{X_2+X_1}{\tau_g(\omega-\gamma \sin2\varphi)} + %\frac{X_2-X_1}{\tau_l(\omega+\gamma \sin2\varphi)}\right]
%\end{equation}
%where
\begin{equation}\label{eq:def_X1}
X_1\equiv\omega (\hat p_g+ \hat p_l)-\gamma \sin(2\phi)(\hat p_g- \hat p_l) 
\end{equation}
and
\begin{equation}\label{eq:def_X2}
X_2\equiv-\gamma \sin(2\phi) (\hat p_g+ \hat p_l)+ \omega (\hat p_g - \hat p_l).
\end{equation}
satisfy a much simpler system of linear first order differential 
equations:
%The first equation, for $X_1$, is trivial whose solution is a constant. The second %equation, for $X_2$, can be solved using an integrating factor. We rewrite it in %the form
\begin{equation}\label{eq:diff_eq}
\frac{dX_1}{d\varphi} = 0 \quad\text{and}\quad
\frac{dX_2}{d\varphi} + A_+(\varphi)X_2 = -A_-(\varphi)X_1,
\end{equation}
where,
\begin{equation}\label{eq:A}
A_\pm(\varphi) = \dfrac{\tau_l(\omega+\gamma \sin2\varphi) \pm \tau_g(\omega-\gamma \sin2\varphi)}{\tau_g\tau_l(\omega^2-\gamma^2\sin^2(2\varphi))}.
\end{equation}
The first equation, for $X_1$, is trivial whose solution is a constant. The second equation, for $X_2$, can be solved exactly with the formal solution:
%using an integrating factor
%Making use of an integrating factor, we can write the formal solution as
\begin{multline}\label{eq:X2}
X_2(\varphi) = \exp\left[-\int_{-\frac{\pi}{2}}^\varphi A_+(\varphi')d\varphi'\right]\cdot\int_{-\frac{\pi}{2}}^\varphi-A_-(\varphi')X_1\exp\left[\int_{-\frac{\pi}{2}}^{\varphi'} A_+(\varphi'')d\varphi''\right]d\varphi' \\
+ C\exp\left[-\int_{-\frac{\pi}{2}}^\varphi A_+(\varphi')d\varphi'\right],
\end{multline}
The arbitrary constant of integration $C$ is determined from the condition of 
a periodic solution, i.e. $X_2(\varphi)=X_2(\varphi+\pi)$ so that we find:
%Furthermore, $X_1$ is a constant and $C$ is the integration constant determined by %the periodicity of the function:
\begin{equation}\label{eq:integration_constant}
C=\frac{\int_{-\frac{\pi}{2}}^{\frac{\pi}{2}}-A_-(\varphi')X_1\exp\left[\int_{-\frac{\pi}{2}}^{\varphi'}A_+(\varphi'')d\varphi''\right]d\varphi'}{\exp\left[\int_{-\frac{\pi}{2}}^{\frac{\pi}{2}}A_+(\varphi')d\varphi'\right]-1}.
\end{equation}
Reversing the relations (\ref{eq:def_X1}) and (\ref{eq:def_X2})
in terms of $\hat p_g$  and $\hat p_l$ and inserting the results into 
(\ref{eq:criterion_fourier}), we obtain a simpler expression for 
the stability criterion:
%In order to simplify (\ref{eq:criterion_fourier}) further, we change of variables 
%By applying this variable change to the unperturbed master equation in %(\ref{eq:ME_transform}) it can be shown that $X_1(\varphi)$ is independent of %$\varphi$, i.e. a constant (see \ref{sec:unperturbed_problem}). Furthermore, 
%by applying this change of variables to (\ref{eq:criterion_fourier}) we get %the expression
\begin{equation}\label{eq:criterion2}
\frac{d}{dt}\langle \ln J(t) \rangle \stackrel{t\rightarrow \infty}{=} \frac{4\gamma\omega\int_{-\pi/2}^{\pi/2}\cos^2\varphi\left(\omega^2-\gamma^2\sin^22\varphi\right)^{-1}X_2(\varphi)\;d\varphi}{X_1\pi(\omega^2-\gamma^2)^{-1/2}+\gamma\int_{-\pi/2}^{\pi/2}\sin2\varphi\left(\omega^2-\gamma^2\sin^22\varphi\right)^{-1}X_2(\varphi)\;d\varphi}.
\end{equation}

%\subsection{Asymptotic approximation [{\color{blue}\emph{To be completed}}]}
\subsection{Large $\tau_l$, $\tau_g$ limit} 
\label{sec:asymptotic_approx}

We can simplify further and solve (\ref{eq:criterion_fourier}) 
in some particular but relevant cases. In the large $\tau_l$, $\tau_g$ limit, but keeping the 
ratio  $\tau_l/\tau_g$ constant, the asymptotic solution developed in 
\ref{sec:unperturbed_problem} can be integrated  
exactly. We have the two cases:

\medskip

\noindent
{\it 1) Underdamped case: $\gamma \leq \omega$}  

\noindent
In this limit the exponential terms inside (\ref{eq:X2}) and 
(\ref{eq:integration_constant}) reduce to unity and the solution simplifies to a constant:
\begin{equation}
X_2(\varphi)=C=-\frac{\int_{-\frac{\pi}{2}}^{\frac{\pi}{2}}
A_-(\varphi')X_1 d\varphi'}
{\int_{-\frac{\pi}{2}}^{\frac{\pi}{2}}A_+(\varphi')d\varphi'}
=\frac{\tau_g-\tau_l}{\tau_g+\tau_l}.
\end{equation}
where the last line results from integration over $\varphi'$.
Inserting this result into (\ref{eq:criterion2}), we obtain 
after integration the stability criterion:
\begin{equation}\label{eq:eta}
\eta = 2\gamma\frac{\tau_g-\tau_l}{\tau_g+\tau_l}.
\end{equation}
After some straightforward algebra, we find 
the angle probability distribution:
\begin{equation}
\hat p_g=\frac{\tau_g X_1}{\omega -\sin(2\varphi) \gamma}  
\quad {\rm and} \quad \hat p_l=\frac{\tau_l X_1}{\omega +\sin(2\varphi) \gamma}.  
\end{equation}
We immediately see that these expressions are $\mathcal{PT}$ symmetric when 
$\tau_l=\tau_g$.
Normalizing these expressions to unity, we fix the constant to
\begin{equation}
X_1=\frac{\sqrt{\omega^2-\gamma^2}}{2\pi(\tau_g + \tau_l)}.
\end{equation}

\medskip
\noindent
{\it 2) Overdamped case: $\gamma \geq \omega$} 

\noindent
In this case $X_2$ is also a constant except at the 
singularities $\varphi_0$, which are the zeroes of 
$\omega= \pm \gamma \sin(2\varphi_0)$.
Some of these singularities are dominant in the sense that 
their weights are much greater than than those of the others. Let us focus our analysis 
on the interval $\varphi \in  ]-\pi/2,\pi/2]$. In order to determine 
their importance, we notice that around these singularities, the 
master equation in (\ref{eq:ME_transform}) decouples so that 
we can make the approximation:
\begin{equation}
\label{eq:sing}
\left[(\omega\mp\gamma\sin2\varphi)\frac{d}{d\varphi}\mp 2\gamma \cos 2\varphi +1/\tau_{g/l} \right]\hat p_{g/l}(\varphi) \simeq 0,
\end{equation}
We can expand this equation locally around the singularity 
$\varphi_0$ to obtain more simply:
\begin{equation}
\label{eq:sing2}
\left[\mp 2\gamma \cos 2\varphi_0 
\left((\varphi -\varphi_0)\frac{d}{d\varphi}+1\right)  +1/\tau_{g/l} \right]\hat p_{g/l}(\varphi) \simeq 0,
\end{equation}
The solution is then
\begin{equation}
\hat p_{g/l}(\varphi) \sim 
|\varphi -\varphi_0|^{-1\mp (2\gamma \cos(2\varphi_0) \tau_{g/l})^{-1}}
\end{equation}
so that the dominant singularities appear for the highest negative power 
fixed by the condition: $\pm \cos(2\varphi_0) >0$. 
After a little algebra, we find that these singularities impose 
the {\it ratio locking} condition: 
\begin{equation}
\tan\varphi_{g/l}=\omega\frac{x_{g/l}}{v_{g/l}}=\pm\frac{\gamma}{\omega}-\sqrt{\left(\frac{\gamma}{\omega}\right)^2-1}.
\end{equation}
Therefore we deduce for the probability distribution:
\begin{equation}
\hat p_g=\frac{\tau_g}{\tau_g +\tau_l}\delta(\varphi -\varphi_g)  
\quad {\rm and} \quad \hat p_l=\frac{\tau_l }{\tau_g +\tau_l}  
\delta(\varphi- \varphi_l).
\end{equation}
The relative weight between the two probabilities is determined by noticing 
that the total probabilities for the gain and loss states should satisfy 
$P_g(t)/P_l(t)\stackrel{t\rightarrow \infty}{=}\tau_g/\tau_l$. 
Contrary to the underdamped case, these probability expressions are not $\mathcal{PT}$ symmetric, no matter the parameter values chosen.
Inserting these last results into 
(\ref{eq:criterion_fourier}), we find after integration 
\begin{equation}
\eta =2\gamma\left(\frac{\tau_g-\tau_l}{\tau_g+\tau_l}+ \sqrt{1-\left(\omega/\gamma\right)^2} \right).
\end{equation}
Note that a similar reasoning can be done in the small $\tau_l$, $\tau_g$ limit keeping the ratio  $\tau_l/\tau_g$ constant. In that case, we would recover (\ref{eq:eta}) provided $\tau_g$ is not too far from $\tau_l$.

%\subsection{Ratio locking [{\color{blue}\emph{To be completed}}]}\label{sec:ratio_locking}

\section{Eigenvalues for the first moment}\label{sec:first_moment_evalue}

The eigenvalues of the system of equations in (\ref{eq:DS}) is calculated directly from the matrix
\begin{equation}
\bm M =
 \begin{pmatrix}
  -1/\tau_l & 1 & 1/\tau_g & 0 \\
  -\omega^2 & -1/\tau_l-2\gamma & 0 & 1/\tau_g \\
  1/\tau_l  & 0  & -1/\tau_g & 1  \\
  0 & 1/\tau_l & -\omega^2 & -1/\tau_g+2\gamma 
 \end{pmatrix}
\end{equation}
whose four eigenvalues are given by the expression
\begin{equation}\label{eq:eigenvalues}
\lambda_i = -\frac{\tau_g+\tau_l\pm\sqrt{A\pm4\tau_g\tau_l\sqrt{B}}}{2\tau_g\tau_l}\quad i=1,2,3,4,
\end{equation}
where
\begin{align}
A&=\tau_l^2+2\tau_g\tau_l(1-2\gamma\tau_l) +\tau_g^2(1+4\gamma\tau_l+\tau_l^2(8\gamma^2-4\omega^2))\notag\\
B&=\tau_l^2(\gamma^2-\omega^2)+(\tau_g+2\gamma\tau_g\tau_l)^2(\gamma^2-\omega^2)-2\tau_g\tau_l(\gamma^2+2\gamma^3\tau_l+\omega^2-2\gamma\omega^2\tau_l).\notag
\end{align}
In the symmetric case, where $\tau_l=\tau_g=\tau$, the system is never stable since there exists at least one eigenvalue that is positive for any combinations of $\gamma$ and $\tau$. Precisely,
\begin{equation}\label{eq:lambda_symmetric}
\lambda_4 = \frac{-1+\sqrt{1+\tau^2(2\gamma^2-1)+2\sqrt{\gamma^2(\gamma^2-1)
\tau^4-\tau^2}}}{\tau}.
\end{equation}
If we assume that $\lambda_4=\alpha + i\beta$ and make the substitution $\sqrt{\gamma^2(\gamma^2-1)\tau^4-\tau^2}=\alpha'+i\beta'$, then by solving for $\alpha$, it is straightforward to show that $Re\lambda_4>0$ for every positive $\gamma$ and $\tau$. Note that $\gamma$ and $\tau$ are real so that $\alpha'$ and $\beta'$ cannot be non-zero at the same time. 

%Finally, we show that $\lim_{\tau_l\rightarrow\infty}\lambda_3 = \gamma -1/\tau_g$.

%\todo{something missing here}

\section{Exact solution for $\omega=0$} \label{omega0}

The probability equations associated to (\ref{vel}) are:
\begin{eqnarray}
(\partial_t \pm 2\gamma \partial_v v)p_{g/l}(v,t)=\mp \left(\frac{p_g(v,t)}{\tau_g}-\frac{p_e(v,t)}{\tau_l}\right).
\end{eqnarray}
We start with the initial condition: $p_{g/l}(v,t=0)=p_{g/l}\delta(v-v_0)$
and use the variable change: $r=\ln(v/v_0)/(2\gamma)$. We make also the transformation
\begin{eqnarray}
p_{g/l}(v,t)=\left(\frac{\tau_{g/l}}{\tau_{l/g}}\right)^{1/4}\frac{1}{v}\exp\left(-\frac{r+t}{2\tau_g}-\frac{t-r}{2\tau_l}\right)
\psi_{g/l}(r,t).
\end{eqnarray}
The new function solves the 1+1 Dirac equation with complex mass $m=i/\sqrt{\tau_l \tau_g}$ :
\begin{eqnarray}\label{dirac}
(\partial_t \pm \partial_r )\psi_{g/l}(r,t)=\psi_{g/l}(r,t)/\sqrt{\tau_g \tau_l}.
\end{eqnarray}
Using the Laplace and Fourier transforms:
\begin{eqnarray}
\psi_{g/l}(k,s)=\int_0^\infty dt \int_{-\infty}^\infty dr e^{-st-ikr}\psi_{g/l}(r,t),
\end{eqnarray}
we solve (\ref{dirac}) to obtain the solution
\begin{eqnarray}
\psi_{g/l}(r,t)&=&\int_{i\delta-\infty}^{i\delta +\infty} \frac{ds}{2\pi i}\int_{-\infty}^\infty \frac{dr}{2\pi}\exp({st+ikr})
\nonumber \\
&&
\times
\frac{\left(\frac{\tau_{g/l}}{\tau_{l/g}}\right)^{1/4}\frac{p_{l/g}}{\sqrt{\tau_g \tau_l}}+
(s \mp i k)\left(\frac{\tau_{l/g}}{\tau_{g/l}}\right)^{1/4}p_{g/l}}
{s^2+k^2-1/\sqrt{\tau_g \tau_l}},
\end{eqnarray}
where $\delta$ has to be chosen such as to leave the poles on the left in the complex plane. Note the dispersion relation 
$\omega= is =\pm \sqrt{k^2-1/\sqrt{\tau_g \tau_l}}$ of the relativistic particle with negative mass. After calculation we obtain for 
the probability:
\begin{eqnarray}
p_{g/l}(v,t)&=&\frac{1}{v}\exp\left(-\frac{r+t}{2\tau_g}-\frac{t-r}{2\tau_l}\right)
\nonumber \\
&&
\times (p_{g/l} \partial_{t\mp r} +\frac{p_{l/g}}{2\tau_{l/g}})1^+(t^2-r^2)I_0\left(\sqrt{\frac{t^2-r^2}{\tau_l \tau_g}}\right),
\end{eqnarray}
where we define the modified Bessel function $I_0(x)=\sum_{n=0}^{\infty} \frac{x^{2n}}{(n!)^2 2^{2n}}\stackrel{x\rightarrow \infty}{=}
\exp(x)/\sqrt{2\pi x}$. 
Quite generally, we find a distribution confined inside the light cone $r=\pm t$. In the short time limit, we recover (\ref{shorttime}).
In the asymptotic limit of large time $t\gg r$ we find instead a normal 
distribution
\begin{eqnarray}
p_{g/l}(v,t\rightarrow \infty)=\frac{\tau_{g/l}}{\tau_g +\tau_l}
\frac{1}{\sqrt{2\pi}\sigma(t)v}
\exp[-(2\gamma)^2(r-r_0(t))^2/2\sigma^2(t)]
\end{eqnarray}
where we recover the average trajectory 
$r_0=\langle \ln(v/v_0) \rangle /(2\gamma)$ and the variance, both given by (\ref{av}).
The large time solution allows us to conclude that the increase of the 
velocity logarithm occurs on average for 
$\tau_g > \tau_l$ and decrease in the contrary case. On the other hand, the stochastic aspect induces always a normal diffusion of this quantity with a relative variance that scales like $1/\sqrt{t}$. Again let us note 
that the asymptotic limit is independent of the chosen weight $p_{g/l}$ at $t=0$ and depends only on the initial speed. 
Finally, we derive also the useful asymptotic  characteristic function:
\begin{eqnarray}\label{charfun}
\ln \langle e^{\alpha r} \rangle \stackrel{t \rightarrow \infty}{=}
\left(\sqrt{\left(\frac{1}{\tau_g}+\frac{1}{\tau_l}\right)^2+4\alpha^2-4\alpha\left(\frac{1}{\tau_g}-\frac{1}{\tau_l}\right)}
-\frac{1}{\tau_g}-\frac{1}{\tau_l}\right)\frac{t}{2}. 
%\geq 0  \quad \quad  \alpha \in R
\end{eqnarray}
It shows that the distribution follows the central limit theorem for large time. This is verified by showing that all cumulants 
scale like $t$. In the particular case where $\alpha =2\gamma n$, 
we deduce the $n$-moment average $\langle v^n \rangle$ from which we recover 
the stability criterion (\ref{n}).

\section*{References}
\bibliographystyle{unsrt}
\bibliography{references}

\end{document}